\documentclass[11pt]{article}
\usepackage{old-arrows}
\usepackage[scaled=.9]{helvet}
\usepackage{charter}
\usepackage{XoohmG}
 \newmdenv[linecolor=blue!60!teal,
 linewidth=1,roundcorner=2pt,backgroundcolor=yellow!3!red!1,
 innerleftmargin=5pt,innerrightmargin=5pt,leftmargin=0pc,rightmargin=0pt,
 fontcolor=blue!60!black,
 ]{bBox}
\usepackage{cite}

\let\fnS=\footnotesize
\def\sfT{{\textsf{T}}\!}
\def\tP{{}^\wtd\!}
\newcommand{\FF}[2][n]{F^{\sss(#1)}_{#2}}

 \SfTitles
 \allowdisplaybreaks
 \numberwithin{equation}{section}
 
 \omitBackreferences

\begin{document}

\thispagestyle{empty}
\setcounter{page}{0}
\vglue5mm
\begin{center}
{\LARGE\sf\bfseries\boldmath
  Beyond Algebraic Superstring Compactification
}
\vspace{2mm}

{\sf\bfseries Tristan H\"{u}bsch$^\ddag$}\\*[1mm]
{\small\it
      Department of Physics \&\ Astronomy,
      Howard University, Washington, DC 20059, USA}\\[-1mm]
 {\tt thubsch@howard.edu}
\vspace{2mm}

{\sf\bfseries ABSTRACT}\\[3mm]
\parbox{152mm}{\addtolength{\baselineskip}{-3pt}\parindent=2pc\noindent
Superstring compactifications have been vigorously studied for over four decades, and have flourished, involving an active iterative feedback between physics and (complex) algebraic geometry. This led to an unprecedented wealth of constructions, virtually all of which are ``purely'' algebraic. Recent developments however indicate many more possibilities to be afforded by including certain generalizations that, at first glance at least, are not algebraic --- yet fit remarkably well within an overall mirror-symmetric framework and are surprisingly amenable to standard computational analysis upon certain mild but systematic modifications.
}

\vspace{5mm}
\begin{minipage}{.75\hsize}\small
  \baselineskip=10pt plus1pt minus 1pt
  \renewcommand{\cftbeforetoctitleskip}{\smallskipamount}
  \renewcommand{\cftaftertoctitle}{\vskip-5pt}
  \tocloftpagestyle{empty}
  \setcounter{tocdepth}{3} 
  \tableofcontents
\end{minipage}
\end{center}

\section{Introduction, Rationale and Summary}
\label{s:IRS}
For over a century, classical geometry of spacetime has been identified with solutions to Einstein's field equations, 
 $R_{\mu\nu}\<=\frac{8\pi G}{c^4}
  \big[\delta_{\mu\nu}^{\rho\sigma}{-}\frac12g_{\mu\nu}g^{\rho\sigma}\big]
  T_{\rho\sigma}$, given here in the ``trace-reversed form'':
 $R_{\mu\nu}\<=R_{\mu\rho\nu}{}^\rho$ is the Ricci tensor, and
 $T_{\mu\nu}$ the energy-momentum density tensor of matter present in the region of interest.
 The geometry of {\em\/empty spacetime\/} ($T_{\mu\nu}\!=\!0$) is thus by definition Ricci-flat. 
 This general qualification is remarkably persistent, through higher-dimensional models and including string theory, where it insures quantum stability to lowest order in string tension~\cite{rF79a, rF79b, Polchinski:1998rq, Polchinski:1998rr}, and then also emerges in the full, oriented loop-space reformulation~\cite{rFrGaZu86, rBowRaj87, rBowRaj87a, rBowRaj87b, Oh:1987sq, rHHRR-sDiffS1, Pilch:1987eb, rBowRaj88, Bowick:1988nj, Bowick:1990wt, rBeast2}.
 Modulo additive total derivatives, $R_{\mu\nu}$ is the 1st Chern class of the underlying spacetime, which links to topology and algebraic geometry and identifies stringy spacetimes, $\sZ$, by the hallmark ``Calabi--Yau'' condition, $c_1(\sZ)\!=\!0$~\cite{rBeast2}.

 Models within this string theory framework that come close to reproducing the observed world include ``Calabi--Yau compactifications''~\cite{rCHSW, rBeast2} and their various generalizations, described by additional conditions involving higher Chern classes of $\sZ$ and of a variety of additional structures, typically defined over $\sZ$. For example, the ``Hull--Strominger system''~\cite{Hull:1985zy, Strominger:1986uh, Hull:1986kz, Becker:2009df, Dasgupta:1999ss} allows (geometric) torsion in $\sZ$ and with additional gauge-field fluxes (possibly deforming its tangent bundle to higher-rank stable bundles) leads to more general, non-K{\"a}hler compactifications~\cite{rBBGDS2,rBBDG1}.\footnote{The underlying worldsheet formulation of string theory pairs its ubiquitous antisymmetric abelian gauge 2-form with the metric into the complex combination, $B_{\m\n}{+}ig_{\m\n}$, at least if the target spacetime (or a factor thereof) permits a locally defined complex structure. The so {\em\/complexified\/} metric structure on the target space a priori permits the metric itself to degenerate at certain locations if balanced by a non-degenerate $B_{\m\n}$ there. This is the general mechanism that extends (analytically continues) GLSMs from a (familiar) ``geometric'' target space to various other ``non-geometric'' descriptions~\cite{rPhases, rMP0}; subsequent analyses have then described those phases using some {\em\/different geometry,} such as {\em\/stratifies pseudo-manifolds\/} turning up in~\cite{rAGM04} and discussed further in~\cite{Hubsch:2002st}, and perhaps even more general notions of geometry may be implied by the inclusion of more general ``defects''~\cite{rB-IS+ST, McNamara:2019rup}.} A survey of considered spacetime geometries aiming also for the observed 3+1-dimensional asymptotically de~Sitter spacetime as given in~\cite{Berglund:2022qsb} indicates some of the complexities of our ultimate goal.
 
Since the pioneering works~\cite{Kasner:1921Fin, rCF-BH}, all algebraic constructions start from some well-understood {\em\/ambient\/} space, $X$, within which the desired geometries are found as solutions of systems of algebraic equations, $Z_f\!:=\!\{x\!\in\!X,~f(x)\!=\!0\}$. To this end, $X$ is typically chosen to have some degree of (quasi) homogeneity~\cite{rBeast2}, which corresponds to a gauge symmetry in the underlying super-conformal quantum field theory on the worldsheet swept by the superstring. With at least $(0,2)$-supersymmetry and
 $U(1)^n\!\too{\sss\>\text{susy}}U(1;\IC)^n\!=\!(\IC^*)^n$ abelian gauge symmetry on the worldsheet, this defines the broad class of {\em\/gauged linear sigma models\/} (GLSMs)~\cite{rPhases, rMP0, Distler:1993mk, Sharpe:2024dcd}, and {\em\/billions\/}\footnote{Many of the half a billion convex reflexive polytopes~\cite{Kreuzer:2000xy, wKS-CY} admit distinct triangulations, leading to distinct toric varieties and so to distinct deformation families of hypersurfaces in them. Each of these constructions admits additional variations by deforming the tangent bundle of the hypersurface and extending it via additional line-bundles, ultimately leading to an astounding combinatorial wealth~\cite{Constantin:2018xkj}.} of catalogued Calabi--Yau 3-folds~\cite{rH-CY0, rGHCYCI, rCYCI1, rCLS-WCP4, rKreSka95, SKARKE_1996, rKreSka98, Kreuzer:2000xy, wKS-CY}. These constructions are vigorously studied by physicists (mostly string theorists) and mathematicians (mostly algebraic geometers) alike, owing to the inherent use of complex algebraic and toric geometry~\cite{rGrHa, rD-TV, rF-TV, rGE-CCAG, rCLS-TV}.
 
The impressive body of work related to these complex algebraic toric geometry constructions notwithstanding, the original string theory requirements mentioned in the 2nd paragraph in fact mandate neither supersymmetry nor complex structure in the ``target spacetime,'' $\sZ$.
 Indeed, neither of these structures is guaranteed in the requisite worldsheet $(0,1)$-supersymmetric formulation; see~\cite{rUDSS01, rUDSS02, rUDSS04, rUDSS09} and much more recently~\cite{Kusuki:2024gtq} for another foundational aspect. In this sense, target spacetime supersymmetry and complex structure (at least in the compact factors of $\sZ$) provide a robust and rigid framework affording a very high degree of computability, a vast array of lampposts to illuminate the landscape.%
 \footnote{This reminds of Danilov's description of ``{\em frigid toric crystals\/}''~\cite[\SS~0.6, p.\,100]{rD-TV}.} %
\ %
However, such models can at most be regarded as an approximation to describing the real world, where supersymmetry is at most a broken symmetry.
 It is then very gratifying to find that the transposition mirror model construction~\cite{rBH, rBH-LGO+EG, Krawitz:2009aa, rLB-MirrBH} does extend to the more general spaces described in Refs.~\cite{rBH-Fm, rBH-gB, Berglund:2022dgb, Berglund:2024zuz, Hubsch:2025sph} --- although many of the details (and possible limitations!) of such extensions still remain to be determined, especially from the symplectic geometry vantage point.

The primary objective of this article then indeed is to serve as a descriptive and motivational rallying exposition of these extensions, calling attention both to their established features and to some of the key open questions. This aims to catalyze further research towards a more robust (foundational, rigorous) and comprehensive understanding.
To that end,
 \SS\,\ref{s:FmFam} presents and explores an infinite sequence  of GLSMs the ground states of which form double deformation families of Calabi--Yau hypersurfaces in Hirzebruch scrolls; their analysis is facilitated by being realized both in the (complex-algebraic) toric geometry framework and as generalized complete intersections in products of projective spaces. 
 \SS\,\ref{s:TM} presents how the transposition mirror construction extends to these models and provides for a vast combinatorial array of multiple mirrors.
 \SS\,\ref{s:CC} concludes by highlighting some open questions and concerns brought about by extending the construction and computational framework to include embedding Calabi--Yau manifolds in non-Fano algebraic varieties as well as non-algebraic torus manifolds, which will hopefully serve as a challenge for future development.

\section{A Showcasing Deformation Family}
\label{s:FmFam}
Expanding on earlier work~\cite{rBH-Fm, rBH-gB, Berglund:2022dgb, Berglund:2024zuz}, I focus on a particular infinite sequence of constructions corresponding to Calabi--Yau hypersurfaces in Hirzebruch scrolls, the rich mathematical history of the latter~\cite{rH-Fm, rGrHa, rF-TV, rGE-CCAG, rCLS-TV} affording their study from multiple different points of view, including their realization within GLSM models~\cite{rMP0} --- which is where we start.

\subsection{From GLSM to Toric}
\label{s:FmFam1}
Let us begin with listing a collection of (chiral superfield) variables and their $U(1)^2$-charges:
\begin{equation}
    \begin{array}{r|r|cccccc@{\,}l}
  &x_0 &x_1 &x_2 &x_3 &x_4 &x_5 &x_6 \\[-1pt] \cmidrule[.8pt]{1-8}
 Q^1 &-4 &~~1 & 1 & 1 & 1 & 0 & 0 &
    \multirow2*{~\smash{$\bigg\}U(1)^2$}}\\ 
 Q^2 &m{-}2 &-m & 0 & 0 & 0 & 1 & 1 \\ 
  \end{array}
 \label{e:X1-6}
\end{equation}
The fact that $Q^a(x_0)$ cancels (for each $a\!=\!1,2$) the sum of charges of the other variables guarantees that the $U(1)^2$ gauge anomalies cancel, and also that the product $x_0(x_1\cdots x_6)$ is $U(1)^2$-invariant and so admissible in the superpotential. $x_0$ may be thought of as a quantum field theory generalization of a Lagrange multiplier~\cite{rUDSS08, rUDSS09, rChaSM, rSingS, rPhases}. Owing to its ubiquitous importance in the algebraic geometry of the deformations of the superpotential,
 $\Pi{x}\<\coeq(\prod_i x_i)$ has been dubbed the ``fundamental monomial''~\cite{rHY-SL2}; see also~\cite{rPeriods1}.

The choice of the superpotential provides part of the defining equations for the ``ground state variety,'' by definition of the {\em\/potential energy\/}:
\begin{equation}
  U(x_i,\s_a)
  = \sum_i\Big|\pd{W}{x_i}\Big|^2
   +\frac{e^2}2\sum_a\,\Big|\sum\nolimits_iQ^a_i|x_i|^2\!-\!r_a\Big|^2
   +2\sum_{a,b,i} \7\s_a\s_b Q^a_iQ^b_i|x_i|^2,
 \label{e:U}
\end{equation}
where $W(x)=x_0\,f(x)$ is the general form of the superpotential,
$Q^a_i\!\coeq\!Q^a(x_i)$, and $\s^a$ is the (twisted-chiral superfield) variable corresponding to the $a^\text{th}$ $U(1)$ gauge group~\cite{rPhases, rMP0}.
Supersymmetric ground states are zeros of~\eqref{e:U}, and the combined vanishing of the 1st (``F'') and 2nd (``D'') terms in~\eqref{e:U}, associate distinct $r_a$-regions with the distinct $U(1)^2$-orbits in the $x_i$-space. In particular, $r_a\!>\!0$ provides the ``geometric phase'' wherein the location
\begin{equation}
   \{x_1=\cdots=x_4=0\}\cup\{x_5=0=x_6\}
 \label{e:Ex}
\end{equation}
is excluded; for details of the GLSM motivations and the other phases/options, see~\cite{rPhases, rMP0, rBH-gB}. The remaining $X$-space is projectivized by the supersymmetry-complexified gauge symmetry, $(\IC^*)^2$, specifying the Hirzebruch scroll,
 $\FF[4]m$. Indeed, $x_1,\cdots,x_6$ in~\eqref{e:X1-6} will be identified with its Cox coordinates, and the $Q^a$-rows specify its Mori vectors and Chern class~\cite{rCLS-TV},
\begin{equation}
  c(\FF[4]m)=\prod_{i=1}^6\Big(1+\sum_{a=1}^2Q^a_iJ_a\Big),\quad
  c_1(\FF[4]m)=4J_1+(2{-}m)J_2,
\end{equation}
so $\deg(f)\<=\deg(x_1\cdots x_6)\<=\binom4{2{-}m}$ also guarantees the zero-locus, $Z_f\subset\FF[4]m$, to be Ricci-flat.

\paragraph{The Superpotential}
 The general form of $W(x)\<=x_0\,f(x)$ implies at the minimum
 of~\eqref{e:U}:
\begin{equation}
  \Big|\pd{W}{x_i}\Big|^2=0:\quad
  f(x)\!=\!0 ~~~\&~~~ x_0\,f'_i(x)\!=\!0.
 \label{e:F}
\end{equation}
This is highly singular for the simple choice $f(x)=\Pi{x}$, which we aim to smooth by including other charge-$\binom{4}{2-m}$ monomials, which are of the form:
\begin{equation}
   x_1^{~k}\, \big(x_2\<\oplus x_3\<\oplus x_4\big)^{4-k}\,
              \bigg\{\Big(x_5\<\oplus x_6\Big)^{2+(k-1)m}
              _{k\geqslant1 \<\vee m\leqslant2}~~\text{or}~~
              \Big(\frac1{x_5}\<\oplus \frac1{x_6}\Big)^{m-2}
              _{k=0,~ m\geqslant3} \bigg\},\quad
   k=0,\dots,4.
 \label{e:MI}
\end{equation}
This at once shows that for $m\<\geqslant3$, the monomials~\eqref{e:MI} are either all proportional to $x_1$ or involve negative powers of $x_5,x_6$. Standard practice in {\em\/both\/} quantum field theory and algebraic geometry is to omit rational monomials, which crucially hobbles the intended deformations: Restricting to $k\<=1,\cdots4$, insures that all {\em\/regular\/} polynomial choices,
\begin{equation}
  f(x) = x_1\cdot \Big(\cC\coeq
                       x_1^{~k}(x_2\<\oplus x_3\<\oplus x_4)^{3-k}
                            (x_5\<\oplus x_6)^{2+km},~k\<=0,\cdots,3\Big),
 \label{e:NpS}
\end{equation}
necessarily factorize so that the zero-locus reduces: 
$Z_f\<=\{x_1\<=0\}\cup\{\cC\<=0\}$, and the singular locus,\break
 $Z_f^\sharp\<=\{x_1\<=0\}\cap\{\cC\<=0\}$ is a Calabi--Yau 2-fold.
 The 3-fold $Z_f$ is Tyurin degenerate~\cite{Berglund:2022dgb} and deemed ``unsmoothable'' as there are no regular monomials to render~\eqref{e:NpS} transverse.

In turn, setting $k\<=0$ in~\eqref{e:MI} identifies monomials that are $x_1$-independent and so can smooth~\eqref{e:NpS} by deforming it away from $\{x_1{=}0\}\subset Z_f^\sharp$.
\begin{figure}[htb]
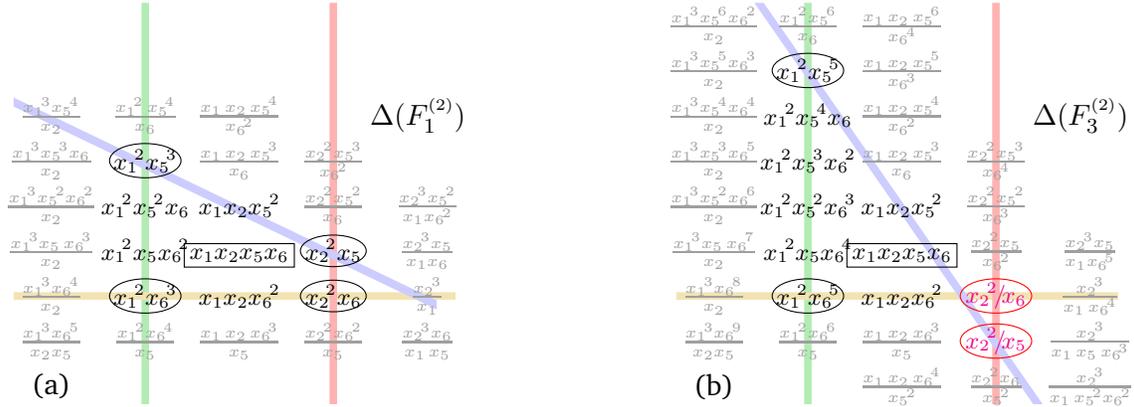

 \centering
  \TikZ{[xscale=1.25, yscale=.6]
    \path[use as bounding box](-2.3,-3.2)--(2.2,5.3);
    \draw[Green!30, line width=1mm](-1,5.5)--++(0,-8.9);
    \draw[yellow!50!brown!40, line width=1mm](-2.4,-1)--++(4.7,0);
    \draw[red!30, line width=1mm](1,5.5)--++(0,-8.9);
    \fill[blue!20]
        (-2.4,3.2)--++(0,.2)--++(4.5,-4.5)--++(0,-.2);
    \path(1.9,3)node{$\D(\FF[2]1)$};
    \path(-1, 2)node{\fnS$x_1\!^2x_5\!^3$};
     \draw(-1, 2)circle(3.8mm);
    \path(-1, 1)node{\fnS$x_1\!^2x_5\!^2x_6$};
    \path(-1, 0)node{\fnS$x_1\!^2x_5x_6\!^2$};
    \path(-1,-1)node{\fnS$x_1\!^2x_6\!^3$};
     \draw(-1,-1)circle(3.8mm);
    \path( 0, 1)node{\fnS$x_1x_2x_5\!^2$};
    \path( 0,-.1)node[rectangle, draw=black, inner sep=2pt]
        {\fnS$x_1x_2x_5x_6$};
    \path( 0,-1)node{\fnS$x_1x_2x_6\!^2$};
    \path( 1, 0)node{\fnS$x_2\!^2x_5$};
     \draw( 1, 0)circle(3.5mm);
    \path( 1,-1)node{\fnS$x_2\!^2x_6$};
     \draw( 1,-1)circle(3.5mm);
    \path(-2,-3)node{(a)};
\color{gray!75}
    \path(-2, 3)node{\fnS$\frac{x_1\!^3x_5\!^4}{x_2}$};
    \path(-2, 2)node{\fnS$\frac{x_1\!^3x_5\!^3x_6}{x_2}$};
    \path(-2, 1)node{\fnS$\frac{x_1\!^3x_5\!^2x_6\!^2}{x_2}$};
    \path(-2, 0)node{\fnS$\frac{x_1\!^3x_5\,x_6\!^3}{x_2}$};
    \path(-2,-1)node{\fnS$\frac{x_1\!^3x_6\!^4}{x_2}$};
    \path(-2,-2)node{\fnS$\frac{x_1\!^3x_6\!^5}{x_2x_5}$};
    \path(-1,-2)node{\fnS$\frac{x_1\!^2x_6\!^4}{x_5}$};
    \path( 0,-2)node{\fnS$\frac{x_1\,x_2\,x_6\!^3}{x_5}$};
    \path( 1,-2)node{\fnS$\frac{x_2\!^2x_6\!^2}{x_5}$};
    \path( 2,-2)node{\fnS$\frac{x_2\!^3x_6}{x_1\,x_5}$};
    \path( 2,-1)node{\fnS$\frac{x_2\!^3}{x_1}$};
    \path( 2, 0)node{\fnS$\frac{x_2\!^3x_5}{x_1x_6}$};
    \path( 2, 1)node{\fnS$\frac{x_2\!^3x_5\!^2}{x_1x_6\!^2}$};
    \path( 1, 1)node{\fnS$\frac{x_2\!^2x_5\!^2}{x_6}$};
    \path( 1, 2)node{\fnS$\frac{x_2\!^2x_5\!^3}{x_6\!^2}$};
    \path( 0, 2)node{\fnS$\frac{x_1\,x_2\,x_5\!^3}{x_6}$};
    \path( 0, 3)node{\fnS$\frac{x_1\,x_2\,x_5\!^4}{x_6\!^2}$};
    \path( -1, 3)node{\fnS$\frac{x_1\!^2\,x_5\!^4}{x_6}$};
            }
\qquad\qquad\qquad\qquad
  \TikZ{[xscale=1.25, yscale=.6]
    \path[use as bounding box](-2.3,-3.2)--(2.2,5.3);
    \draw[Green!30, line width=1mm](-1,5.5)--++(0,-8.9);
    \draw[yellow!50!brown!40, line width=1mm](-2.4,-1)--++(4.7,0);
    \draw[red!30, line width=1mm](1,5.5)--++(0,-8.9);
    \fill[blue!20]
        (-{1.57},5.5)--++(.1,0)--++(8.9/3,-8.9)--++(-.1,0);
    \path(1.9,3)node{$\D(\FF[2]3)$};
    \path(-1, 4)node{\fnS$x_1\!^2x_5\!^5$};
     \draw(-1, 4)circle(3.8mm);
    \path(-1, 3)node{\fnS$x_1\!^2x_5\!^4x_6$};
    \path(-1, 2)node{\fnS$x_1\!^2x_5\!^3x_6\!^2$};
    \path(-1, 1)node{\fnS$x_1\!^2x_5\!^2x_6\!^3$};
    \path(-1, 0)node{\fnS$x_1\!^2x_5x_6\!^4$};
    \path(-1,-1)node{\fnS$x_1\!^2x_6\!^5$};
     \draw(-1,-1)circle(3.8mm);
    \path( 0, 1)node{\fnS$x_1x_2x_5\!^2$};
    \path( 0,-.1)node[rectangle, draw=black, inner sep=2pt]
        {\fnS$x_1x_2x_5x_6$};
    \path( 0,-1)node{\fnS$x_1x_2x_6\!^2$};
    \path[magenta]( 1,-1)node{\fnS$x_2\!^2\!/\!x_6$};
     \draw[red]( 1,-1)circle(3.8mm);
    \path[magenta]( 1,-2)node{\fnS$x_2\!^2\!/\!x_5$};
     \draw[red]( 1,-2)circle(3.8mm);
    \path(-2,-3)node{(b)};
\color{gray!75}
    \path(-2, 5)node{\fnS$\frac{x_1\!^3x_5\!^6x_6\!^2}{x_2}$};
    \path(-2, 4)node{\fnS$\frac{x_1\!^3x_5\!^5x_6\!^3}{x_2}$};
    \path(-2, 3)node{\fnS$\frac{x_1\!^3x_5\!^4x_6\!^4}{x_2}$};
    \path(-2, 2)node{\fnS$\frac{x_1\!^3x_5\!^3x_6\!^5}{x_2}$};
    \path(-2, 1)node{\fnS$\frac{x_1\!^3x_5\!^2x_6\!^6}{x_2}$};
    \path(-2, 0)node{\fnS$\frac{x_1\!^3x_5\,x_6\!^7}{x_2}$};
    \path(-2,-1)node{\fnS$\frac{x_1\!^3x_6\!^8}{x_2}$};
    \path(-2,-2)node{\fnS$\frac{x_1\!^3x_6\!^9}{x_2x_5}$};
    \path(-1,-2)node{\fnS$\frac{x_1\!^2x_6\!^6}{x_5}$};
    \path( 0,-2)node{\fnS$\frac{x_1\,x_2\,x_6\!^3}{x_5}$};
    \path( 0,-3)node{\fnS$\frac{x_1\,x_2\,x_6\!^4}{x_5\!^2}$};
    \path( 1,-3)node{\fnS$\frac{x_2\!^2x_6}{x_5\!^2}$};
    \path( 2,-3)node{\fnS$\frac{x_2\!^3}{x_1\,x_5\!^2x_6\!^2}$};
    \path( 2,-2)node{\fnS$\frac{x_2\!^3}{x_1\,x_5\,x_6\!^3}$};
    \path( 2,-1)node{\fnS$\frac{x_2\!^3}{x_1\,x_6\!^4}$};
    \path( 2, 0)node{\fnS$\frac{x_2\!^3x_5}{x_1x_6\!^5}$};
    \path( 1, 0)node{\fnS$\frac{x_2\!^2x_5}{x_6\!^2}$};
    \path( 1, 1)node{\fnS$\frac{x_2\!^2x_5\!^2}{x_6\!^3}$};
    \path( 1, 2)node{\fnS$\frac{x_2\!^2x_5\!^3}{x_6\!^4}$};
    \path( 0, 2)node{\fnS$\frac{x_1\,x_2\,x_5\!^3}{x_6}$};
    \path( 0, 3)node{\fnS$\frac{x_1\,x_2\,x_5\!^4}{x_6\!^2}$};
    \path( 0, 4)node{\fnS$\frac{x_1\,x_2\,x_5\!^5}{x_6\!^3}$};
    \path( 0, 5)node{\fnS$\frac{x_1\,x_2\,x_5\!^6}{x_6\!^4}$};
    \path(-1, 5)node{\fnS$\frac{x_1\!^2x_5\!^6}{x_6}$};
            }
 \caption{Some of the $c_1(\FF[2]m)$-degree monomials plotted to indicate the ``strips'' discussed in the text; the fundamental monomial, $x_1x_2x_5x_6$ is boxed}
 \label{f:F1F3M}
\end{figure}
To this end, compare the $c_1$-degree monomials for $\FF[2]1$ and $\FF[2]3$, shown in Figure~\ref{f:F1F3M} for the 2-dimensional surfaces (adjusting to $\deg[c_1(\FF[2]m)]\<=\binom2{2-m}$ and omitting $x_3,x_4$) for simplicity. The plots make evident that:
\begin{enumerate}[itemsep=0pt, topsep=1pt, labelsep=1.17pc]

 \item Monomials independent of a particular variable occur along a straight-line ``stripe'' (hyperplane in higher dimensions). Therefore, each ``stripe'' is a suitable multiple of a single $x_j$-derivative, $(\vd_j\Pi{x})$:
\begin{equation}
  \begin{array}{@{}r@{~~}cccc@{}}
 k\in\ZZ
 &x_1\text{-indep.} &x_2\text{-indep.}
 &x_5\text{-indep.} &x_6\text{-indep.} \\ \toprule
 \textbf{gen.:}
 &(x_2 x_5\!^k x_6\!^{-k-m})\vd_1
 &(x_1 x_5^k x_6^{m-k})\vd_2
 &(x_1^k x_2^{-k} x_6^{1+km})\vd_5
 &(x_1^k x_2^{-k} x_5^{1+km})\vd_6 \\[4pt]
 \textbf{stripe:}
 &x_2^2 x_5^{1+k} x_6^{1-k-m}
 &x_1^2 x_5^{1+k} x_6^{1-k+m}
 &x_1^{1+k} x_2^{1-k} x_6^{2+km}
 &x_1^{1+k} x_2^{1-k} x_5^{2+km}
\end{array}
 \label{e:stripes}
\end{equation}
 Since $\vd_j^2\Pi{x}\<=0$, each ``stripe'' acts as a boundary
 --- for {\em\/that\/} $\vd_j$-deformation.

 \item ``Cornerstone'' monomials at the intersection of two ``stripes'' are independent of {\em\/two\/} variables; this hierarchy extends straightforwardly in higher dimensions. The tabulation~\eqref{e:stripes} makes it clear that:
 ({\small\bf1})~There is no $x_1$- and $x_2$-independent monomial.
 ({\small\bf2})~There is an $x_5$- and $x_6$-independent monomial only for $m\<=1,2$, $x_2\!^2/x_1$ and $x_2\!^2$, respectively.
 ({\small\bf3})~The ``cornerstone'' monomials are $x_1\!^2 x_5\!^{2+m}$, 
 $x_1\!^2 x_6\!^{2+m}$, $x_2\!^2 x_5\!^{2-m}$ and $x_2\!^2 x_6\!^{2-m}$,
 and are circled in the plots in Figure~\ref{f:F1F3M}.

 \item The above shows that deforming the fundamental monomial equips the system of anticanonical monomials with the hierarchical structure of a poset:
\begin{equation}
   \vC{\TikZ{[thick, xscale=1.5,
              every node/.style={inner sep=0, outer sep=2}]
     \path[use as bounding box](-3,0)--(3,2.2);
     \node[outer sep=3](0) at(0,0) {$\Pi{x}$};
     \node(1)  at(-3,1) {\fnS$x_2^2 x_5^{1+k} x_6^{1-k-m}$};
     \node(5)  at(-1,1) {\fnS$x_1^{1+k} x_2^{1-k} x_6^{2+km}$};
     \node(6)  at( 1,1) {\fnS$x_1^{1+k} x_2^{1-k} x_5^{2+km}$};
     \node(2)  at( 3,1) {\fnS$x_1^2 x_5^{1+k} x_6^{1-k+m}$};
     \node(15) at(-3,2) {\fnS$x_2\!^2 x_6\!^{2-m}$};
     \node(16) at(-1,2) {\fnS$x_2\!^2 x_5\!^{2-m}$};
     \node(25) at( 1,2) {\fnS$x_1\!^2 x_6\!^{2+m}$};
     \node(26) at( 3,2) {\fnS$x_1\!^2 x_5\!^{2+m}$};
     \draw[-stealth](0)--node[below left]{\fnS$\vd_1$}(1);
     \draw[-stealth](5)--(15);
     \draw[-stealth](6)--(16);
     \draw[densely dotted, very thick, -stealth]
         (0)--node[above right=-1pt]{\fnS$\vd_5$}(5);
     \draw[densely dotted, very thick, -stealth](1)--(15);
     \draw[densely dotted, very thick, -stealth](2)--(25);
     \draw[densely dashed, very thick, -stealth]
         (0)--node[above left=-1pt]{\fnS$\vd_6$}(6);
     \draw[densely dashed, very thick, -stealth](1)--(16);
     \draw[densely dashed, very thick, -stealth](2)--(26);
     \draw[double, -stealth](0)--node[below right=1pt]{\fnS$\vd_2$}(2);
     \draw[double, -stealth](5)--(25);
     \draw[double, -stealth](6)--(26);
      }}
 \label{e:posetM}
\end{equation}

\end{enumerate}\smallskip
Being reachable by a simple ({\em\/first\/}) derivative from the fundamental monomial, $\Pi{x}$, monomials on each $x_i$-independent ``stripe'' are at (deformation) distance of 1 from $\Pi{x}$, with the direction of the respective deformations indicated by the Euclidean lattice normals:
\begin{equation}
\vC{\TikZ{[ultra thick]\path[use as bounding box](-1.5,-1)--(1.5,1);
          \foreach\x in{-1,...,1}\foreach\y in{-1,...,1}
           \fill[gray](\x,\y)circle(.5mm);
          \corner{(0,0)}{0}{45}{.6}{purple}
          \corner{(0,0)}{45}{180}{.5}{teal}
          \corner{(0,0)}{180}{270}{.5}{Green}
          \corner{(0,0)}{270}{360}{.5}{orange}
          \draw[blue, -stealth](0,0)--(1,1);
           \path[blue](1,1)node[right]{$\vd_6$};
          \draw[red, -stealth](0,0)--(1,0);
           \path[red](1,0)node[right]{$\vd_1$};
          \draw[yellow!30!brown, -stealth](0,0)--(0,-1);
           \path[brown](0,-1)node[right]{$\vd_5$};
          \draw[Green, -stealth](0,0)--(-1,0);
           \path[Green](-1,0)node[left]{$\vd_2$};
          \filldraw[fill=white, thick](0,0)circle(.5mm);
          \path(.2,.6)node[left]{$\S(\FF[2]1)$};
          \path(-.67,-.9)node{(a)};
            }}
\qquad\qquad
\vC{\TikZ{[ultra thick]\path[use as bounding box](-1.5,-1)--(3.5,1);
          \foreach\x in{-1,...,3}\foreach\y in{-1,...,1}
           \fill[gray](\x,\y)circle(.5mm);
          \corner{(0,0)}{0}{atan(1/3)}{.8}{purple}
          \corner{(0,0)}{atan(1/3)}{180}{.5}{teal}
          \corner{(0,0)}{180}{270}{.5}{Green}
          \corner{(0,0)}{270}{360}{.5}{orange}
          \draw[blue, -stealth](0,0)--(3,1);
           \path[blue](3,1)node[right]{$\vd_6$};
          \draw[red, -stealth](0,0)--(1,0);
           \path[red](1,0)node[right]{$\vd_1$};
          \draw[yellow!30!brown, -stealth](0,0)--(0,-1);
           \path[brown](0,-1)node[right]{$\vd_5$};
          \draw[Green, -stealth](0,0)--(-1,0);
           \path[Green](-1,0)node[left]{$\vd_2$};
          \filldraw[fill=white, thick](0,0)circle(.5mm);
          \path(.2,.6)node[left]{$\S(\FF[2]3)$};
          \path(-.67,-.9)node{(b)};
            }}
 \label{e:F1F3N}
\end{equation}
Each 2-dimensional cone enclosed between two of these consecutive directions corresponds to a (circled) corner monomial in Figure~\ref{f:F1F3M}:
 $\sfa(\vd_1,\vd_6)\iff x_2\!^2x_5$, $\sfa(\vd_2,\vd_5)\iff x_1\!^2x_5\!^3$, etc. The so-constructed {\em\/fan\/} of cones, $\S(\FF[2]{m})$, in fact specifies the Hirzebruch scroll $\FF[2]m$ as a toric variety~\cite{rF-TV, rGE-CCAG, rCLS-TV}.
 
 The fans~\eqref{e:F1F3N} have a natural dimension-ranked poset structure generated by the inclusion of cones in the boundary of one-higher dimensional cones, and isomorphic to~\eqref{e:posetM}: In~\eqref{e:F1F3N}, 
 the central 0-cone is 
 within the boundary of the 1-cone $\vd_1$, 
 which is within the boundary of the 2-cone $\sfa(\vd_1,\vd_6)$.
This chain of relations is strictly ({\em\/inclusion-reversing\/}) dual to the corresponding statements regarding the monomials in Figure~\ref{f:F1F3M}; for example:
\begin{equation}
  \begin{array}{@{}r@{:\quad}ccccc@{}}
 \eqref{e:F1F3N} &
 0 \text{~(center)} &\subset&
 \text{boundary of }\sfa(\vd_1) &\subset&
 \text{boundary of }\sfa(\vd_1,\vd_6)\\
 \text{Figure~\ref{f:F1F3M}} &
  \text{all monom's} &\supset&
   x_1\text{-indep.\ monom's} &\supset&
   x_1,x_6\text{-indep.\ monom's} \\
\end{array}
\end{equation}
 Also of note is the fact that the 2-dimensional cones
 $\sfa\!(\vd_1,\vd_2)$ and $\sfa\!(\vd_5,\vd_6)$
 do not belong to either of the two (posets) fans~\eqref{e:F1F3N}; this defines the so-called Stanley-Reisner ideal among the linear vector (sub)spaces generated by $\vd_i$~\cite{rCLS-TV}. Dually in Figure~\ref{f:F1F3M}, the two vertical ``stripes'' (monomials without $x_1$ and without $x_2$, respectively) never intersect, and the horizontal ($x_5$-omitting)  ``stripe'' intersects the slanted ($x_6$-omitting) ``stripe'' either outside the distance-1 convex polygon enclosing the universal monomial, $\Pi{x}$, or at a non-lattice location, $(\frac2m,-1)$ for $\FF[2]m$, where that distance-1 enclosing polygon self-intersects; this defines the so-called ``irrelevant'' ideal among the multiplicative ring of monomials~\cite{rCLS-TV}.

\paragraph{The Transpolar Operation}
The ``stripe''-wise dual operation used above to map the monomial systems in Figure~\ref{f:F1F3M} to the fans~\eqref{e:F1F3N} is a simple version of the {\em\/transpolar\/} operation (denoted by ``$\,^\wtd$''; see \SS\,\ref{s:TM}) defined more formally in Ref.~\cite{rBH-gB, Berglund:2022dgb, Berglund:2024zuz}. It implements the standard {\em\/polar\/} operation of algebraic toric geometry~\cite{rF-TV, rGE-CCAG, rCLS-TV} for each (convex subset of each) face of a polytope, then reassembles the resulting elements using the canonical inclusion-reversing nature of any duality.
 Moreover, the same iterative operation also works perfectly in reverse:
 The Euclidean normal to the $[\vd_5,\vd_1]$ ``stripe'' in~\eqref{e:F1F3N}
  is $(1,-1)$ and indicates $x_2\!^2x_6$ for $\FF[2]1$
  and $x_2\!^2\!/\!x_6$ for $\FF[2]1$;
 the normal to $[\vd_1,\vd_6]$ in~(\ref{e:F1F3N},\,a)
  is $(1,0)$ and indicates $x_2\!^2x_5$ for $\FF[2]1$, while
 the normal to $[\vd_1,\vd_6]$ in~(\ref{e:F1F3N},\,b)
  is $(1,-2)$ and indicates $x_2\!^2\!/\!x_5$ for $\FF[2]3$, and so on.
 The key distinction in $\D(\FF[2]1)$ between $m\<=0,1,2$ (convex and flat) and $m\<\geqslant3$ (self-intersecting, i.e., flip-folded) cases reflects the fact that the integral hull (lattice enclosure) of $\S(\FF[2]m)$ is convex for $m\<=0,1,2$ but non-convex for $m\<\geqslant3$.

This transpolar operation is markedly {\em\/unlike\/} the standard polar operation:\footnote{It was highly amusing to learn that many active researchers in algebraic toric geometry, while always citing the standard (global) polar operation, in practice often use variants of the (local) ``stripe''-wise dual (dubbed ``transpolar''~\cite{rBH-gB}) operation --- the original invention of which is ``lost in the mists of time''; I thank Hal Schenck for communicating these tidbits.}
 Defined {\em\/globally\/} over the entire polyhedral body at once,
 when starting from $\S(\FF[2]3)$ in~\eqref{e:F1F3N}, the standard polar operation automatically replaces it with the {\em\/convex hull,\/} which obscures the generator
 $\vd_1$; the so-chosen monomial set in Figure~\ref{f:F1F3M}\,(b) stops at the intersection of the slanted and horizontal ``stripes,'' never reaching the right-hand side vertical (red) stripe of monomials (which are indeed all rational).
 When starting from $\D(\FF[2]3)$ in Figure~\ref{f:F1F3M}\,(b) including also the right-hand side circled monomials, the standard polar maps only to $\{\vd_1,\vd_2\}$. When omitting the right-hand side circled monomials, the right-most remaining monomials include $\Pi{x}$ and define a distance-0 ``stripe,'' which cannot define a Euclidean normal: Not only is the standard polar operation not involutive, but it is ill-defined for $\FF{m}$ when $m\<\geqslant3$ --- which are non-Fano.
 
Comparing the two monomial plots in Figure~\ref{f:F1F3M} and their transpolar images in~\eqref{e:F1F3N} reveals several key features:
Whereas the simple, flat fans in~\eqref{e:F1F3N} evidently subdivide the (enveloping) polygons that span them, the arrangements of ``stripes'' present a nontrivial distinction: The left-hand side $\D(\FF[2]1)$ is a plain, flat trapezoid and has a simple subdividing fan.
 In turn, $\D(\FF[2]3)$ is subdivided by a fan-like structure (corresponding to a toric space) if it is understood to be a flip-folded and multi-layered {\em\/multitope,} which spans and is subdivided by a {\em\/multifan\/}~\cite{rM-MFans, Masuda:2000aa, rHM-MFs, Masuda:2006aa, rHM-EG+MF, rH-EG+MFs2, Nishimura:2006vs, Ishida:2013ab}; see also~\cite{Davis:1991uz, Ishida:2013aa, buchstaber2014toric, Jang:2023aa}. We will return to this in \SS\,\ref{s:TM}.

\begin{corl}\label{C:GLSM-tP}
For a list of (chiral superfield) variables and their $U(1;\IC)^n$-charges as in~\eqref{e:X1-6}, the most general superpotential, $W$ in~\eqref{e:U}, is an $X_0$-multiple of a deformation the fundamental monomial, $\Pi{x}$.
 The lattice of all candidate monomials, $\D(X)$, has hyperplanes at 1-derivative distance from $\Pi{x}$, the deformation directions of which span the (multi)\,fan, $\S(X)$, that corresponds to the underlying toric (ambient) space wherein the ground states minimize the potential~\eqref{e:U}.
 
 The so-defined $\D(X)\<{\too{\text{\tiny\,GLSM\,}}}\,\S(X)$ mapping a priori and by definition selects the {\em\/transpolar\/} extension~\cite{rBH-gB, Berglund:2022dgb, Berglund:2024zuz} of the standard polar operation~\cite{rF-TV, rGE-CCAG, rCLS-TV}.
\end{corl}
\begin{remk}\label{R:GLSM-tP}
Curbing the $\Pi{x}$-deformations (as in Figure~\ref{f:F1F3M}) to only the regular monomials (with only non-negative powers) in the selected (Cox) variables~\eqref{e:X1-6} restricts the transpolar
 $\D(X)\<{\too{\text{\tiny\,GLSM\,}}}\,\S(X)$ mapping so it agrees with the standard polar operation --- but only provided ``fractional,'' non-lattice locations are also included: In $\D(\FF[2]3)$ in Figure~\ref{f:F1F3M}, this is the $(\frac23,-1)$ intersection of the slanted and horizontal ``stripes,'' which corresponds to $\sqrt[3]{x_1x_2\!^5}$ --- also beyond the standard practice in complex-algebraic toric geometry!%
 \footnote{The radical monomial, $\sqrt[3]{x_1x_2\!^5}$, reminds of similar factors that were found to play the role of ``twisted vacua'' in Landau--Ginzburg orbifolds~\cite{rRes}. The self-crossing region in the complete Newton multitope, $\D(\FF{m})$, for $m\<\geqslant2$ corresponds to monomials of the form $\big[{\oplus_{i=2}^n}\big(\frac{x_i}{x_1}\big){}^{\frac1m}\big]{}^2{\cdot}x_1 \cdot \big({\oplus_{j=2}^n}x_j\big)^{n-1}$, which depend only on the fiber-coordinates in
 $\FF{m}\<=\ssK[{r||c}{\IP^n&1\\ \IP^1&m}]$,
but where the first, radical factor has degree
 $\binom02\<=\deg[\rd y_0\,\rd y_1]$ of the volume-from of the base-$\IP^1$.
This is indeed the ``hallmark'' quality whereby radical monomials enable representing ``non-polynomial'' deformations~\cite{rRes}. For $m\<\geqslant3$, such radical deformations are still proportional to a positive power of $x_1$ and so could not smooth the Tyurin-degenerate regular Calabi--Yau hypersurfaces.}
The oft-required ``regularity'' (non-negative powers) depends on the choice of variables, and is not as unequivocal as may be expected; see~\eqref{e:dx3}--\eqref{e:CI2TV3} below.
\end{remk}
\begin{remk}\label{R:GLSM-CI}
The foregoing extends to systems of multiple defining constraints: complete intersections, certain non-complete intersections and higher-rank constraints defined by exact sequences of direct sums of line-bundles are straightforward (though tedious)~\cite{rBeast2}. Extensions to non-abelian GLSM gauge groups are also possible~\cite{Hori:2016txh, Ruan:2017Non}, also beyond our present scope.
\end{remk}

\begin{remk}\label{R:RegRat}
Regarding the system of anticanonical monomials (as in Figure~\ref{f:F1F3M}) as candidate $\vd_i$-directional \eqref{e:F1F3N} deformations of the fundamental monomial, $\Pi{x}$, shows that the distance-1 rational monomial deformations are all, a priori, ``accessible'' --- by deforming in only one
 $\vd_i$-direction at a time: In Figure~\ref{f:F1F3M}\,(b) plot, the distance-1 monomials in the right-hand side vertical (red) ``stripe'' form the boundary for $\vd_1$-deformations. Among them,
 $x_2\!^2\!/\!x_5$ is indeed inaccessible by $\vd_5$-deformations and
 $x_2\!^2\!/\!x_6$ by $\vd_6$-deformations;
 they are accessible as $x_2/x_5\!^2x_6$- and $x_2/x_5x_6\!^2$-multiples of
 $\vd_1$-deformations wherein $x_2,x_5,x_6$ do not vary.
The standard {\em\/polar\/} operation~\cite{rF-TV, rGE-CCAG, rCLS-TV} imposes such limitations {\em\/all at once\/} and is in this sense ``{\em\/global.}''
 In contradistinction, the ``stripe''-wise {\em\/transpolar\/} operation~\cite{rBH-gB, Berglund:2022dgb} is ``{\em\/local.}''
\end{remk}
\begin{remk}\label{R:WhyL}
Conversely, the principal reason for including the rational sections such as $x_2\!^2\!/\!x_5$ and $x_2\!^2\!/\!x_6$ in Figure~\ref{f:F1F3M}\,(b) is that the segment of ``stripe'' they form is transpolar to the $\vd_1$-deformation, which is a generator of the fan~(\ref{e:F1F3N},\,b) that encodes $\FF[2]3$; see also Figure~\ref{f:2F3tPm}, below.
 Omitting the $\vd_1$-generator from~(\ref{e:F1F3N},\,b) and its transpolar (rational) monomials encodes $\IP^2_{(1:1:3)}\neq\FF[2]3$.
\end{remk}

\subsection{From Toric to Generalized Intersections}
\label{s:FmFam2}
While ultimately interested in Calabi--Yau hypersurfaces in 4-folds, we note that the two particular Hirzebruch scrolls defined by the fans~\eqref{e:F1F3N} and equipped with the anticanonical sections plotted in Figure~\ref{f:F1F3M} are well known to be {\em\/diffeomorphic\/} to each other; they are the same real, smooth manifold equipped however with discretely distinct complex structures.
 Nevertheless, there exists an explicit, continuous $\e$-deformation family of hypersurfaces, extending Hirzebruch's original~\cite{rH-Fm}:
\begin{equation}
  \big\{ p_\e(x,y) = x_0 y_0\!^3 +x_1 y_1\!^3 +\e\,x_2 y_0\!^2 y_1 = 0,~~
  \e\<\in\IC \big\} =
  \K[{r||c}{\IP^2_x&1\\[1pt] \IP^1_y&3}]
 \label{e:HFme}
\end{equation}
At the center, $\e=0$, is Hirzebruch's original scroll,
 $\{p_0(x,y)\<=0\}\subset\IP^2_x{\times}\IP^1_y$, herein denoted $\FF[2]3$; extending $\IP^2_x\leadsto\IP^n_x$ defines $\FF{m}$ as the zero-locus of the same $p_\e(x,y)$. Hirzebruch scrolls have a hallmark submanifold, the directrix~\cite{rGrHa} with a maximally negative self-intersection ($-m$), and which can actually be specified as an explicit hypersurface even in the realization~\eqref{e:HFme}, using a recent construction~\cite{rgCICY1}; see also~\cite{rBH-Fm, rGG-gCI}, and~\cite{Berglund:2022dgb} for an explicit algorithm. The degree of the directrix (with respect to $\IP^2_x{\times}\IP^1_y$) must be
 $\binom{~~1}{-3}$, which of course cannot be holomorphic on all of
 $\IP^2_x{\times}\IP^1_y$. However, the equivalence class
\begin{equation}
 \Fs(x,y) \coeq
  \Big[\Big(\frac{x_0}{y_1\!^3}{-}\frac{x_1}{y_0\!^3}\Big) 
       +\lambda\frac{p_0(x,y)}{(y_0y_1)^3}\Big]
 =\bigg\{ \begin{array}{@{}l@{~\text{where}~}l@{~\text{with}~}l}
           +2\frac{x_0}{y_1\!^3} &y_1{\neq}0, &\lambda=+1;\\[2mm]
           -2\frac{x_1}{y_0\!^3} &y_0{\neq}0, &\lambda=-1;
          \end{array}
 \label{e:dx3}
\end{equation}
is indeed holomorphic on
 $\FF[2]3\<\coeq\{p_0(x,y)\<=0\}\subset\IP^2_x{\times}\IP^1_y$:
Each of the two representatives is manifestly holomorphic over their respective (one-point-punctured) part of $\IP^1_y$, and their difference on the overlap is a finite multiple of $p_0(x,y)$, which vanishes on $\FF[2]3$ by definition.
 Furthermore, the constant-Jacobian variable change
\begin{equation}
  \underbrace{(x_0,x_1,x_2;y_0,y_1) = (p_0,\Fs,x_2;y_0,y_1)}
            _{\det[J]\,=\,4} ~~\too{p_0=0}~~
  \begin{array}{r|cccc}
     &x_1{=}\Fs &x_2 &x_5{=}y_0 &x_6{=}y_1 \\[-1pt] \toprule
 Q^1 &~~~1 & 1 & 0 & 0 \\ 
 Q^2 &-3  & 0 & 1 & 1 \\ 
  \end{array}
 \label{e:CI2TV3}
\end{equation}
is a coordinate-level identification of this hypersurface with the toric rendition of $\FF[2]3$, and where the components of the $(Q^a$ row-wise) Mori vectors are equal to the $\IP^2_x{\times}\IP^1_y$ homogeneity degrees.

Away from the center, when $\e\<\neq0$, the directrix-defining section~\eqref{e:dx3} fails to be holomorphic, but is replaced by two that are holomorphic:
\begin{subequations}
 \label{e:dx2+1}
\begin{alignat}9
 \Fs_1(x,y) \coeq
  \Big[\Big(\frac{x_0 y_0}{y_1\!^3} -\frac{x_1}{y_0\!^2}
               +\e\frac{x_2}{y_1\!^2}\Big) 
       +\lambda\frac{p_\e(x,y)}{y_0\!^2y_1\!^3}\Big],&\quad
  \deg&=\tbinom{~~\,1}{-2};\\
 \Fs_2(x,y) \coeq
  \Big[\Big(\frac{x_0}{y_1} -\frac{x_1 y_1\!^2}{y_0\!^3}
            -\e\frac{x_2}{y_0}\Big) 
       +\lambda\frac{p_\e(x,y)}{y_0\!^3y_1}\Big]&\quad
  \deg&=\tbinom{~~\,1}{-1}.
\end{alignat}
\end{subequations}
The constant-Jacobian change of variables
\begin{equation}
  \underbrace{(x_0,x_1,x_2;y_0,y_1) = (p_\e,\Fs_1,\Fs_2;y_0,y_1)}
            _{\det[J]\,=\,4\e} ~~\too{p_\e=0}~~
  \begin{array}{r|cccc}
     &x_1{=}\Fs_1 &x_2{=}\Fs_2 &x_5{=}y_0 &x_6{=}y_1 \\[-1pt] \toprule
 Q^1 &~~~1 &~~~1 & 0 & 0 \\ 
 Q^2 &-2  &-1 & 1 & 1 \\ 
  \end{array}
 \label{e:CI2TV21}
\end{equation}
maps this $(\e{\neq}0)$-family of hypersurfaces to a ``cousin'' of Hirzebruch's scroll, which one might denote $\FF[2]{\sss(2,1)}$. In the GLSM, the $U(1)^2$ charges~\eqref{e:CI2TV21} are trivially redefined:
\begin{equation}
  \begin{array}{r|cccc}
     &x_1{=}\Fs_1 &x_2{=}\Fs_2 &x_5{=}y_0 &x_6{=}y_1 \\[-1pt] \toprule
       Q^1 &~~~1 & 1 & 0 & 0 \\ 
 Q^2{+}Q^1 &-1   & 0 & 1 & 1 \\ 
  \end{array}
 \label{e:CI2TV10}
\end{equation}
specifying the Mori vectors of the standard $\FF[2]1$ --- consistent with the well known diffeomorphism $\FF[2]m\approx_{\sss\IR}\FF[2]{m-2k}$, for $k\<\in\ZZ$. It is tempting to conjecture~\cite{Hubsch:2025sph} that the Segr\'e-like change of variables
\begin{equation}
  \K[{r||c}{\IP^2&1\\ \IP^1&3}]_{\e\neq0}~~\bigg\{
  \begin{array}{@{}r@{~}l@{}}
  (x_0y_0\!^2,x_1y_1\!^2,x_2y_0y_1;y_0,y_1)
    &= (\x_0,\x_1,\x_2;\h_0,\h_1)\\[1mm]
  x_0 y_0\!^3 +x_1 y_1\!^3 +\e x_2 y_0\!^2 y_1
    &\to (\x_0+\e\x_2)\h_0 +\x_1 \h_1
\end{array}
  \bigg\}~~\K[{r||c}{\IP^2&1\\ \IP^1&1}]
 \label{e:Segre}
\end{equation}
provides for the $\FF[2]3\<{\approx_{\sss\IR}}\FF[2]1$
 {\em\/diffeomorphism\/}: Its Jacobian
 ($\det\frac{\vd(\x;\h)}{\vd(x;y)}=(y_0y_1)^3$) is non-constant and vanishes (the inverse diverges) over the poles of $\IP^1$, where it can be modified by partitions of unity (``bump functions'') to make~\eqref{e:Segre} a smooth but non-biholomorphic diffeomorphism. Although the foregoing discussion for simplicity focuses on simple 2-dimensional showcasing examples, generalizations to higher dimensions are straightforward, as a few examples below will show.
\begin{remk}\label{R:chVars}
The evidently self-consistent and effective use in the above context of sections that are
 equivalence classes of Laurent polynomials in terms of $\IP^1_y$ homogeneous coordinates,
 but regular (Cox) variables in the toric reformulation
 lends support to the concerns about the choice of variables raised in Remark~\ref{R:GLSM-tP}: One may wonder if a better choice of variables might render the Laurent (rational-monomial) deformations such as those encircled  in the right-hand side vertical ``stripe'' of the $\D(\FF[2]3)$ plot in Figure~\ref{f:F1F3M}\,(b); this is revisited in \SS\,\ref{s:CC}.
\end{remk}

\subsection{The Deformation Family Picture}
\label{s:FmFam3}
Following the example in \SS\,\ref{s:FmFam2}, consider now the explicit 3-dimensional {\em\/double\/} deformation family of Calabi--Yau hypersurfaces in Hirzebruch scrolls, themselves defined as hypersurfaces in
$\IP^4_x{\times}\IP^1_y$:
\begin{equation}
  \begin{array}{@{}l@{\quad}l@{\quad}r@{\,=\,}l@{}}
   \text{first:} & p_\e(x,y)=0, &\deg[p_\e]&\binom1m \\[3pt]
   \text{then:}  & q_\e(x,y)=0, &\deg[q_\e]&\binom1{2{-}m} \\
  \end{array}\bigg\}:~~
  \K[{r||c|c}{\IP^4&1&4\\ \IP^1&m&2{-}m}].
 \label{e:gCIXm}
\end{equation}
The so-constructed smooth Calabi--Yau manifolds have $h^{11}\<=2\<=b_2$ and 
 $h^{21}\<=86$ so $b_3\<=174$ and $\chi_E\<={-}168$.
 Introduced and dubbed ``generalized complete intersections''~\cite{rgCICY1}, in such deformation families the second constraint polynomial, $q_\e(x,y)$, is defined as a holomorphic section only on the zero locus of the first, $p_\e(x,y)$ and their choices are correlated. This structure also has a well-defined {\em\/scheme-theoretic\/} formulation over 
 $\IP^4_x{\times}\IP^1_y$~\cite{rGG-gCI} and is amenable to a minor modification of the standard (co)homological algebra computations~\cite{rBH-Fm}. Here we focus on the $\vec\e$-family of first hypersurfaces,
\begin{equation}
  p_\e(x,y) = x_0 y_0\!^5 +x_1 y_1\!^5
   +\e_2x_2 y_0\!^4 y_1 +\e_3x_3 y_0\!^3 y_1\!^2 +\e_4x_4 y_0\!^2 y_1\!^3
    = 0,
 \label{e:4Fme}
\end{equation}
which consist of various 4-dimensional Hirzebruch scrolls~\cite{Berglund:2022dgb, Hubsch:2025sph}.
 Within these we then find correlated, $\deg\<=\binom{~~\:4}{-3}$, Calabi--Yau 3-dimensional hypersurfaces, the defining sections of which are closely related to the directrices~\eqref{e:dx3} and~\eqref{e:dx2+1}.

\paragraph{$\vec\e\<={(0,0,0)}$:}
At the center, the $\{p_0(x,y){=}0\}$ hypersurface has the single,
degree-$\binom{~~1}{-5}$ directrix,
\begin{alignat}9
  \Fs_0(x,y) &=
  \Big[\Big( \frac{x_0}{y_1\!^5} {-}\frac{x_1}{y_0\!^5} \Big)
       ~~\mathrm{mod}~ \frac{p_0(x,y)}{(y_0y_1)^5}\Big],
  \qquad \deg{=}\tbinom{~~1}{-5}.
 \label{e:s5}
\intertext{By the constant-Jacobian variable change \`a la~\eqref{e:CI2TV3}, this identifies $\{p_0(x,y){=}0\}\<=\FF[4]5$:}
 &\begin{array}{r|cccccc}
     &x_1{=}\Fs_1 &x_2 &x_3 &x_4 &x_5{=}y_0 &x_6{=}y_1 \\[-1pt] \toprule
 Q^1 &~~~1 & 1 & 1 & 1 & 0 & 0 \\ 
 Q^2 &-5   & 0 & 0 & 0 & 1 & 1 \\ 
  \end{array}
\intertext{and also defines the correlated family of degree-$\binom{~~4}{-3}$ sections:}
  q_0(x,y) &=
  \big( c^{00}(x)y_0\!^2 +c^{01}(x)y_0y_1 +c^{11}(x)y_1\!^2 \big)
  \<\cdot\Fs_0(x,y),
 \label{e:q0}
\end{alignat}
where $c^{ij}(x)$ provide $3{\times}35\<=105$ cubics that parametrize the deformation family of Calabi--Yau 3-folds in $\FF[4]5$. As evident from the explicit factorization of~\eqref{e:q0}, all such Calabi--Yau 3-folds are Tyurin degenerate, deemed unsmoothable (by regular anticanonical sections), but smoothed by the $k\<=0$, $m\<=5$ rational sections~\eqref{e:MI},
 $(x_2\<\oplus x_3\<\oplus x_4)^4
  \big(\frac1{x_5}\<\oplus \frac1{x_6}\big){}^3$.
 They correspond to the segment of the right-hand side vertical distance-1 ``stripe'' in Figure~\ref{f:F1F3M}\,(b) delimited by its intersections with the horizontal and the slanted ``stripes''~\cite{rBH-gB, Berglund:2022dgb}. Tyurin degeneration is characterized by the singular set
\begin{equation}
  \big\{q_0\<=0\big\}^\sharp =
  \big\{ \Fs_0\<=0 \big\} \cap
  \big\{ c^{00}(x)y_0\!^2 {+}c^{01}(x)y_0y_1 {+}c^{11}(x)y_1\!^2 \<=0 \big\}
  \in \K[{r||c|cc}{\IP^4&1&~~1&3\\ \IP^1&5&-5&2\\}],
\end{equation}
which is a K3 surface, a Calabi--Yau {\em\/matryoshka\/} within the degenerate Calabi--Yau 3-fold $\{q_0(x,y)\<=0\}$.

\paragraph{$\vec\e\<=(1,0,0)$:}
The system is now deformed to:
\begin{alignat}9
  p_1(x,y) &=
  x_0 y_0\!^5 +x_1 y_1\!^5  +x_2 y_0\!^4 y_1,\\
 \To~~
 &\Fs_{1a}(x,y) =
  \Big[\Big( \frac{x_0 y_0}{y_1\!^5} {-}\frac{x_1}{y_0\!^4}
            {+}\frac{x_2}{y_1\!^4} \Big)
       ~~\mathrm{mod}~ \frac{p_1(x,y)}{y_0\!^4\,y_1\!^5}\Big],
  &\qquad \deg&=\tbinom{~~1}{-4}, \label{e:s1a}\\
 &\Fs_{1b}(x,y) =
  \Big[\Big( \frac{x_0}{y_1} {-}\frac{x_1 y_1\!^4}{y_0^5}
             {-}\frac{x_2}{y_0}\Big)
       ~~\mathrm{mod}~ \frac{p_1(x,y)}{y_0\!^5\,y_1} \Big],
  &\qquad \deg&=\tbinom{~~1}{-1}. \label{e:s1b}
\intertext{By the constant-Jacobian variable change \`a la~\eqref{e:CI2TV21}, this identifies $\{p_1{=}0\}\<=\FF[4]{\sss(4100)}$:}
 &\begin{array}{r|cccccc}
     &x_1{=}\Fs_{1a} &x_2{=}\Fs_{1b} &x_3 &x_4 &x_5{=}y_0 &x_6{=}y_1 \\[-1pt] \toprule
 Q^1 &~~~1 &~~~1 & 1 & 1 & 0 & 0 \\ 
 Q^2 &-4   &-1 & 0 & 0 & 1 & 1 \\ 
  \end{array}
\intertext{and also defines the correlated family of degree-$\binom{~~4}{-3}$ sections:}
  q_1(x,y) &=
  \makebox[0mm][l]{$\displaystyle
  c^i_a(x)y_i\<\cdot\Fs_{1a}(x,y)
  +c_b(x)\<\cdot\Big[\frac{\Fs_{1b}(x,y)}{y_1\!^2}
       ~~\mathrm{mod}~ \frac{p_1(x,y)}{y_0\!^5\,y_1\!^3}\Big],$}
 \label{e:q1}
\end{alignat}
where again $c^i_a(x),c_b(x)$ provide $3{\times}35\<=105$ cubics that parametrize the deformation family of Calabi--Yau 3-folds in this ``cousin'' Hirzebruch scroll, $\FF[4]{(4100)}$. Since~\eqref{e:q1} does not factorize, this already deforms from the Tyurin degeneration in~\eqref{e:q0}.
 Indeed, generic choices of $\{q_1\<=0\}$ are expected to be singular at most at isolated points: 
 $q_1\sim (c^i_ay_i){\cdot}\Fs_{1a}+c_b{\cdot}(\Fs_{1b}/y_1\!^2)$ is singular at nine (9) points,
 $\{(c^i_ay_i),\,\Fs_{1a},\,c_b,\,\Fs_{1b}\<=0\}$.

\paragraph{$\vec\e\<=(1,1,0)$:}
The system is now deformed to:
\begin{alignat}9
  p_2(x,y) &=
  x_0 y_0\!^5 +x_1 y_1\!^5  +x_2 y_0\!^4 y_1 +x_3 y_0\!^3 y_1\!^2,\\
 \To~~
 &\Fs_{2a}(x,y) =
  \Big[\Big( \frac{x_0 y_0\!^2}{y_1\!^5} {-}\frac{x_1}{y_0\!^3}
             {+}\frac{x_2 y_0}{y_1\!^4} {+}\frac{x_3}{y_1\!^3} \Big)
       ~~\mathrm{mod}~ \frac{p_2(x,y)}{y_0\!^3\,y_1\!^5}\Big],
  &\qquad \deg&=\tbinom{~~1}{-3}, \label{e:s2a}\\
 &\Fs_{2b}(x,y) =
  \Big[\Big( \frac{x_0}{y_1} {-}\frac{x_1 y_1\!^4}{y_0\!^5} 
             {-}\frac{x_2}{y_0} {-}\frac{x_3 y_1}{y_0\!^2} \Big)
       ~~\mathrm{mod}~ \frac{p_2(x,y)}{y_0\!^5\,y_1}\Big],
  &\qquad \deg&=\tbinom{~~1}{-1}, \label{e:s2b}\\
 &\Fs_{2c}(x,y) =
  \Big[\Big( \frac{x_0 y_0}{y_1\!^2} {-}\frac{x_1 y_1\!^3}{y_0\!^4}
           {+}\frac{x_2}{y_1} {-}\frac{x_3}{y_0} \Big)
       ~~\mathrm{mod}~ \frac{p_2(x,y)}{y_0\!^4\,y_1\!^2}\Big],
  &\qquad \deg&=\tbinom{~~1}{-1}. \label{e:s2c}
\intertext{By the constant-Jacobian variable change \`a la~\eqref{e:CI2TV21}, this identifies $\{p_1{=}0\}\<=\FF[4]{\sss(3110)}$:}
 &\begin{array}{r|cccccc}
     &x_1{=}\Fs_{2a} &x_2{=}\Fs_{2b} &x_3{=}\Fs_{2c} &x_4 &x_5{=}y_0 &x_6{=}y_1 \\[-1pt] \toprule
 Q^1 &~~~1 &~~~1 &~~~1 & 1 & 0 & 0 \\ 
 Q^2 &-3   &-1 &-1 & 0 & 1 & 1 \\ 
  \end{array}
\intertext{and also defines the correlated family of degree-$\binom{~~4}{-3}$ sections:}
  q_2(x,y) &=
  c_a(x)\<\cdot\Fs_{2a}(x,y)
  \makebox[0mm][l]{$\displaystyle
  +c_b(x)\<\cdot\Big[\frac{\Fs_{2b}(x,y)}{y_1\!^2}
       ~~\mathrm{mod}~ \frac{p_2(x,y)}{y_0\!^5\,y_1\!^3}\Big]$}\nonumber\\
  &\mkern24mu
  +c_c(x)\<\cdot\Big[\frac{\Fs_{2c}(x,y)}{y_0\,y_1}
       ~~\mathrm{mod}~ \frac{p_2(x,y)}{y_0\!^5\,y_1\!^3}\Big],
 \label{e:q2}
\end{alignat}
where again $c_a(x),c_b(x),c_c(x)$ provide $3{\times}35\<=105$ cubics that parametrize the deformation family of Calabi--Yau 3-folds in this ``cousin'' Hirzebruch scroll, $\FF[4]{(3110)}$. Owing to the swapped sign of the $x_2$-term, $\Fs_{2b}/y_1\!^2$ and $\Fs_{2c}/y_0y_1$ are independent from each other and are also independent from $\Fs_{2a}$.

\paragraph{$\vec\e\<=(1,1,1)$:}
The system is now deformed to:
\begin{alignat}9
  p_3(x,y) &=
  x_0 y_0\!^5 +x_1 y_1\!^5  +x_2 y_0\!^4 y_1 
  +x_3 y_0\!^3 y_1\!^2 +x_4 y_0\!^2 y_1\!^3,\\
 \To~~
 &\Fs_{3a}(x,y) =
  \Big[\Big( \frac{x_0 y_0\!^3}{y_1^5} {-}\frac{x_1}{y_0\!^2}
            {+}\frac{x_2 y_0\!^2}{y_1\!^4} {+}\frac{x_3 y_0}{y_1\!^3}
             {+}\frac{x_4}{y_1^2} \Big)
       ~\mathrm{mod}~ \frac{p_3(x,y)}{y_0\!^2\,y_1\!^5}\Big],
  &~~ \deg&=\tbinom{~~1}{-2}, \label{e:s3a}\\
 &\Fs_{3b}(x,y) =
  \Big[\Big( \frac{x_0}{y_1} {-}\frac{x_1 y_1\!^4}{y_0\!^5}
           {-}\frac{x_2}{y_0} {-}\frac{x_3 y_1}{y_0\!^2}
            {-}\frac{x_4 y_1\!^2}{y_0\!^3} \Big)
       ~\mathrm{mod}~ \frac{p_3(x,y)}{y_0\!^5\,y_1}\Big],
  &~~ \deg&=\tbinom{~~1}{-1}, \label{e:s3b}\\
 &\Fs_{3c}(x,y) =
  \Big[\Big( \frac{x_0 y_0}{y_1\!^2} {-}\frac{x_1 y_1\!^3}{y_0^4}
            {+}\frac{x_2}{y_1} {-}\frac{x_3}{y_0}
               {-}\frac{x_4 y_1}{y_0\!^2} \Big)
       ~\mathrm{mod}~ \frac{p_3(x,y)}{y_0\!^4\,y_1\!^2}\Big],
  &~~ \deg&=\tbinom{~~1}{-1}, \label{e:s3c}\\
 &\Fs_{3d}(x,y) =
  \Big[\Big( \frac{x_0 y_0\!^2}{y_1\!^3} {-}\frac{x_1 y_1\!^2}{y_0\!^3}
           {+}\frac{x_2 y_0}{y_1\!^2} +\frac{x_3}{y_1}
            {-}\frac{x_4}{y_0}\Big)
       ~\mathrm{mod}~ \frac{p_3(x,y)}{y_0\!^3\,y_1\!^3}\Big],
  &~~ \deg&=\tbinom{~~1}{-1}. \label{e:s3d}
\intertext{By the constant-Jacobian variable change \`a la~\eqref{e:CI2TV21}, this identifies $\{p_1{=}0\}\<=\FF[4]{\sss(2111)}$:}
 &\begin{array}{r|cccccc}
     &x_1{=}\Fs_{3a} &x_2{=}\Fs_{3b} &x_3{=}\Fs_{3c} &x_4{=}\Fs_{3d} &x_5{=}y_0 &x_6{=}y_1 \\[-1pt] \toprule
 Q^1 &~~~1 &~~~1 &~~~1 &~~~1 & 0 & 0 \\ 
 Q^2 &-2   &-1 &-1 &-1 & 1 & 1 \\ 
  \end{array}
\intertext{and also defines the correlated family of degree-$\binom{~~4}{-3}$ sections:}
  q_3(x,y) &=
  c_a(x)\<\cdot\Fs_{3a}(x,y)
  \makebox[0mm][l]{$\displaystyle
  +c_b(x)\<\cdot\Big[\frac{\Fs_{3b}(x,y)}{y_1\!^2}
       ~~\mathrm{mod}~ \frac{p_3(x,y)}{y_0\!^5\,y_1\!^3}\Big]$}\nonumber\\
  &\mkern24mu
  \makebox[0mm][l]{$\displaystyle
  +c_c(x)\<\cdot\Big[\frac{\Fs_{3c}(x,y)}{y_0\,y_1}
       ~~\mathrm{mod}~ \frac{p_3(x,y)}{y_0\!^5\,y_1\!^3}\Big]
  +c_d(x)\<\cdot\Big[\frac{\Fs_{3d}(x,y)}{y_0\!^2}
       ~~\mathrm{mod}~ \frac{p_3(x,y)}{y_0\!^5\,y_1\!^3}\Big],$}
 \label{e:q3}
\end{alignat}
where again $c_a(x),\cdots,c_d(x)$ now provide $4{\times}35\<=140$ cubics that overabundantly paramet\-rize the deformation family of Calabi--Yau 3-folds in this ``cousin'' Hirzebruch scroll, $\FF[4]{(2111)}$. Notice that owing to the swapped signs of the $x_2$- and $x_3$-terms, $\Fs_{3b}/y_1\!^2$, $\Fs_{3c}/y_0y_1$ and $\Fs_{3d}/y_0\!^2$ are independent sections.
\ping

Changing variables as in~\eqref{e:CI2TV3} and~\eqref{e:CI2TV21}, the directrices are identified as Cox coordinates in the toric realization of the sequence of Hirzebruch scrolls:
 \eqref{e:s5} in $\FF[4]{(5000)}$, 
 \eqref{e:s1a}--\eqref{e:s1b} in $\FF[4]{(4100)}$, 
 \eqref{e:s2c}--\eqref{e:s2c} in $\FF[4]{(3110)}$, 
 and~\eqref{e:s3a}--\eqref{e:s3d} in $\FF[4]{(2111)}$.
Here found as smooth hypersurfaces in $\IP^4_x{\times}\IP^1_y$ along the explicit deformation path,
 $\vec\e\<=(0,0,0)\<\leadsto(1,0,0)\<\leadsto(1,1,0)\<\leadsto(1,1,1)$,
they are diffeomorphic by deformation, and provide a textbook example of ``the same real manifold'' with a complex structure that varies~\cite{rK+S-DefT} --- herein, discretely.
 Within each of these, there is an effectively 86-dimensional family of Calabi--Yau hypersurfaces, parametrized by the $\IP^4_x$-cubics indicated in~\eqref{e:q0}, \eqref{e:q2}, \eqref{e:q2} and~\eqref{e:q3}, amongst which only those in $\FF[4]{(5000)}$ {\em\/must\/} be Tyurin degenerate, the other ones (moving leftward in the indicated explicit deformation path) acquire an increasing degree of variation and inevitably become transverse.

The real manifold underlying $\FF[4]{(5000)}$, $\FF[4]{(4100)}$, $\FF[4]{(3110)}$ and $\FF[4]{(2111)}$ of the ``first'' deformation family, $\ssK[{r||c}{\IP^4&1\\ \IP^1&5}]$, is also diffeomorphic to
 $\FF[4]1\<=\FF[4]{(1000)}$~\cite{rWall}.
 Indeed, the Segr\'e-like mapping
\begin{alignat}9
  (x_0y_0\!^4, x_1y_1\!^4, x_2y_0\!^3y_1, x_3y_0\!^2y_1\!^2, x_4y_0y_1\!^3; y_0, y_1)
  &\too{\vs} (\x_0, \x_1, \x_2, \x_3, \x_4, \h_0, \h_1)
 \label{e:Segre7}
\iText[-2mm]{converts}
 x_0 y_0\!^5
 {+}\e_2 x_2 y_0^4 y_1
 {+}\e_3 x_3 y_0\!^3 y_1\!^2
 {+}\e_4 x_4 y_0\!^2 y_1\!^3
 {+}x_1 y_1^5
 &\to \h_0 (\x_0 {+}\e_2\x_2 {+}\e_3\x_3 {+}\e_4\x_4) {+}\x_1\h_1,
\end{alignat}
and so maps
 $\ssK[{r||c}{\IP^4&1\\[2pt] \IP^1&5}]\<\ni\FF[4]5
  \dto\FF[4]1\<\in\ssK[{r||c}{\IP^4&1\\[2pt] \IP^1&1}]$. This of course is {\em\/not\/} a biholomorphism, as the Jacobian of the mapping~\eqref{e:Segre7} is non-constant: $\det[\vs]\<=(y_0y_1)^{10}$ vanishes (and the inverse diverges) over the poles of $\IP^1_y$. This can be smoothed without changing the topology by means of local ``bump-functions,'' thereby realizing the non-holomorphic {\em\/diffeomorphism\/} $\FF[4]5\approx_\IR\FF[4]1$. The so-encountered variants of the Hirzebruch scrolls have their hallmark holomorphic and so irreducible directrices of maximally negative self-intersection:
\begin{enumerate}[itemsep=0pt, topsep=1pt, labelsep=1.17pc]
 \item $\FF[4]{(5000)}$ has~\eqref{e:s5} with self-intersection
\begin{equation}
  [\Fs_0^{-1}\!(0)]^4 =
  \ssK[{r||c|cccc}{\IP^4&1&~~~1&~~~1&~~~1&~~~1\\[2pt] 
                   \IP^1&5&-5&-5&-5&-5}]=-15.
 \label{e:si5-5}
\end{equation}
 \item $\FF[4]{(4100)}$ has~\eqref{e:s1a} with self-intersection
\begin{equation}
  [\Fs_{1a}^{-1}\!(0)]^4 =
  \ssK[{r||c|cccc}{\IP^4&1&~~~1&~~~1&~~~1&~~~1\\[2pt] 
                   \IP^1&5&-4&-4&-4&-4}]=-11.
 \label{e:si5-4}
\end{equation}
 \item $\FF[4]{(3110)}$ has~\eqref{e:s2a} with self-intersection
\begin{equation}
  [\Fs_{2a}^{-1}\!(0)]^4 =
  \ssK[{r||c|cccc}{\IP^4&1&~~~1&~~~1&~~~1&~~~1\\[2pt] 
                   \IP^1&5&-3&-3&-3&-3}]=-7.
 \label{e:si5-3}
\end{equation}
 \item $\FF[4]{(2111)}$ has~\eqref{e:s3a} with self-intersection
\begin{equation}
  [\Fs_{3a}^{-1}\!(0)]^4 =
  \ssK[{r||c|cccc}{\IP^4&1&~~~1&~~~1&~~~1&~~~1\\[2pt] 
                   \IP^1&5&-2&-2&-2&-2}]=-3.
 \label{e:si5-2}
\end{equation}
\end{enumerate}
While identical as real, smooth manifolds, the maximally negative self-intersection numbers~\eqref{e:si5-5}--\eqref{e:si5-2} distinguish the
 $\FF[4]{\bS{m}}$ as complex manifolds. The degree-$\binom{~~1}{-1}$ directrices,~\eqref{e:s1b}, \eqref{e:s2b} and~\eqref{e:s2c} and \eqref{e:s3b}--\eqref{e:s3d} have standard, positive self-intersections equal to $+1$.
 The last of the above-calculated,~\eqref{e:si5-2} is in fact identical to the self-intersection of the hallmark holomorphic and so irreducible directrix of $\FF[4]1$,
\begin{equation}
  [\Fs_0^{\prime\,-1}\!(0)]^4 =
  \ssK[{r||c|cccc}{\IP^4&1&~~~1&~~~1&~~~1&~~~1\\[2pt] 
                   \IP^1&1&-1&-1&-1&-1}]=-3,
\end{equation}
which supports the claim that even as complex manifolds, $\FF[4]{(2111)}\approx_\IC\FF[4]1$.

 Since $\FF[4]1$ is Fano, all its generic Calabi--Yau hypersurfaces are smooth.
\ping

Consider now the $\vec\e$-path within the double deformation family of generalized complete intersections~\eqref{e:gCIXm}:
\begin{equation}
   \vC{\begin{tikzpicture}[thick, yscale=.25, xscale=1.25,
      every node/.style={inner sep=0,outer sep=.5mm}]
     \path[use as bounding box](-.5,-.5)--(11,10);
     \node(K2111) at(1,9) {$\cK^*(\FF[4]{(2111)})$};
     \node(F2111) at(1,5) {$\FF[4]{(2111)}$};
     \node[blue](q3)    at(0,2.5) {$\{q_3{=}0\}$};
     \node(111)   at(1,0) {\llap{$\vec\e=\,$}$(1,1,1)$};
     \node(K3110) at(4,9) {$\cK^*(\FF[4]{(3110)})$};
     \node(F3110) at(4,5) {$\FF[4]{(3110)}$};
     \node[blue](q2)    at(3,2.5) {$\{q_2{=}0\}$};
     \node(110)   at(4,0) {$(1,1,0)$};
     \node(K4100) at(7,9) {$\cK^*(\FF[4]{(4100)})$};
     \node(F4100) at(7,5) {$\FF[4]{(4100)}$};
     \node[blue](q1)    at(6,2.5) {$\{q_1{=}0\}$};
     \node(100)   at(7,0) {$(1,0,0)$};
     \node(K5000) at(10,9) {$\cK^*(\FF[4]{(5000)})$};
     \node(F5000) at(10,5) {$\FF[4]{(5000)}$};
     \node[blue](q0)    at( 9,2.5) {$\{q_0{=}0\}$};
     \node(000)   at(10,0) {$(0,0,0)$};
     \draw[blue, -stealth](K2111)to[out=240, in=90]++(-1.25,-5.5);
     \draw[-stealth](K2111)--(F2111);
     \draw[blue, right hook-stealth]([xshift=1mm]q3.north)--(F2111.west);
     \draw[-stealth](F2111)--(111);
     \draw[blue, -stealth](K3110)to[out=240, in=90]++(-1.25,-5.5);
     \draw[-stealth](K3110)--(F3110);
     \draw[blue, right hook-stealth]([xshift=1mm]q2.north)--(F3110.west);
     \draw[-stealth](F3110)--(110);
     \draw[blue, -stealth](K4100)to[out=240, in=90]++(-1.25,-5.5);
     \draw[-stealth](K4100)--(F4100);
     \draw[blue, right hook-stealth]([xshift=1mm]q1.north)--(F4100.west);
     \draw[-stealth](F4100)--(100);
     \draw[blue, -stealth](K5000)to[out=240, in=90]++(-1.25,-5.5);
     \draw[-stealth](K5000)--(F5000);
     \draw[blue, right hook-stealth]([xshift=1mm]q0.north)--(F5000.west);
     \draw[-stealth](F5000)--(000);
     \draw[-stealth, decorate,
     decoration={snake, amplitude=.4mm, segment length=2mm, post length=1mm}]
     (111)--(110);
     \draw[-stealth, decorate,
     decoration={snake, amplitude=.4mm, segment length=2mm, post length=1mm}]
     (110)--(100);
     \draw[-stealth, decorate,
     decoration={snake, amplitude=.4mm, segment length=2mm, post length=1mm}]
     (100)--(000);
     \draw[blue, dotted, -stealth](q3)--(q2);
     \draw[blue, dotted, -stealth](q2)--(q1);
     \draw[blue, dotted, -stealth](q1)--(q0);
     \draw[dashed, -stealth](K2111)--(K3110);
     \draw[dashed, -stealth](K3110)--(K4100);
     \draw[dashed, -stealth](K4100)--(K5000);
   \end{tikzpicture}}
 \label{e:DDDefo}
\end{equation}
The deformation path in the $\vec\e$-space (bottom row, wavy arrows) carries along:
\begin{enumerate}[itemsep=0pt, topsep=1pt, labelsep=1.17pc]
 \item The sequence of (discrete) deformations among the Hirzebruch scrolls, $\FF[2]{\vec\e}$ (middle row in~\eqref{e:DDDefo}, no arrows drawn to avoid clutter), as discussed above.
 \item The sequence of (likewise discrete) deformations (top row in~\eqref{e:DDDefo}, dashed arrows) among the indicated anticanonical bundles, $\cK^*(\FF[2]{\vec\e})$. This induces corresponding deformations on the anticanonical sections, which provide candidates for:
 \item The sequence of discrete ``leafs'' in the {\em\/double deformations\/} of the defining equations (lower middle row in~\eqref{e:DDDefo}, dotted arrows) among the indicated Calabi--Yau hypersurfaces: the $\vec\e$-dependent anticanonical bundles provide sections for the ``second'' deformation of anticanonical hypersurfaces, $\FF[2]{\vec\e}[c_1]$. 
\end{enumerate}

\medskip
The diagram~\eqref{e:DDDefo} prominently features the correlated triple consisting of:
\begin{enumerate}[itemsep=0pt, topsep=1pt, labelsep=1.17pc]
 \item A well-understood embedding (``ambient'') space, $X$,
       chosen $\FF{m}$ in~\eqref{e:DDDefo}.
 \item The anticanonical bundle, $\cK^*(X)$,
       and a collection of its sections, $f\<\in\G(\cK^*(X))$.
 \item The Calabi--Yau hypersurface --- the zero locus, $Z_f$, of a
       selected anticanonical section.
\end{enumerate}
\begin{remk}\label{R:3}
This entire triple plays a prominent role in the definition and analysis of GLSMs~\cite{rPhases, rMP0} and motivates in part the original proposal of Laurent deformations~\cite{rBH-gB}. It is worth noting that $\cK^*(X)$ is also a Calabi--Yau space, albeit non-compact.
 Also, between the {\em\/non-compact\/} $(n{+}1)$-dimensional Calabi--Yau space, $\cK^*(X)$, and the {\em\/compact\/} $(n{-}1)$-dimensional Calabi--Yau zero locus, $Z_f$, of a section of $\cK^*(X)$ is the embedding space $X$, which is not Calabi--Yau --- but wherein $(X\<\ssm Z_f)$ is an $n$-dimensional {\em\/non-compact\/} Calabi--Yau space~\cite{rTY1, rTY2}. This last fact seems to have been noticed only recently, and has so far not been explored.
\end{remk}

\subsection{Black Sheep in the Deformation Family}
\label{s:FmFamBS}
Exploring the diagram~\eqref{e:DDDefo} now in the horizontal direction and knowing that all variants of the embedding space, $\FF{m}$, are in fact the {\em\/same smooth real manifold,\/} it is vexing to find that
all anticanonical sections~\eqref{e:q0} on the far right of the diagram~\eqref{e:DDDefo} factorize (see~\eqref{e:NpS} for the GLSM/toric rendition), so that their zero loci, $Z_{q_0}$, are Tyurin degenerate Calabi--Yau varieties. They are deemed unsmoothable --- since the deformations of $q_0(x,y)$ are routinely limited to regular monomials (with non-negative powers) in the original variables~\eqref{e:X1-6}. Indeed, omitted in~\eqref{e:NpS} for $m\<\geqslant3$ are the $k\<=0$ monomials~\eqref{e:MI}, which provide:
\begin{equation}
  \text{Laurent deformations}:\quad \d_L \big(f(x)=q_0(x,y)\big) = 
   \bigoplus\nolimits_{k=0}^{m-2} \frac{(x_2\<\oplus x_3\<\oplus x_4)^4}
        {x_5^{~k}\, x_6^{~m-2-k}},
 \label{e:LaurDefo}
\end{equation}
and are analogous to the monomials on the segment of the right-hand side vertical distance-1 ``stripe'' in Figure~\ref{f:F1F3M}\,(b) delimited by its intersections with the horizontal and the slanted ``stripe''~\cite{rBH-gB, Berglund:2022dgb}. Being independent of $x_1\<=\Fs_0$, the deformation~\eqref{e:LaurDefo} moves the zero locus away from the singularity $Z_{f=q_0}^\sharp\<=\{x_1{=}0\}\cap\cC(x)$.
In the simpler notation of \SS\,\ref{s:FmFam1},
direct computation then verifies that the generic combinations of~\eqref{e:MI} and~\eqref{e:LaurDefo} are transverse. For a simple example as in \SS\,\ref{s:FmFam2}, consider the Calabi--Yau hypersurface in the family
$\FF[2]3[c_1]$, defined as the zero locus of
\begin{equation}
  f_{\!L\vec\d\,}(x) = x_1^{~2} x_5^{~5} +x_1^{~2} x_6^{~5}
    +\d_1\frac{x_2\!^2}{x_5} +\d_2\frac{x_2\!^2}{x_6},\quad
  \d_1,\d_2\<\neq0~~\text{and}~~\d_1^{~5}\<\neq\d_2^{~5}.
 \label{e:fd12}
\end{equation}
Direct computation verifies that $f_{\!L\vec\d\,}(x)\<=0\<=\rd f_{\!L\vec\d\,}(x)$ only when $x_1\<=0\<=x_2$, which cannot happen in $\FF[2]m$: The change of variables~\eqref{e:CI2TV3} implies that $(\Fs_0{=}x_1)\<=0\<=x_2$ together with
 $\FF[2]3=\{p_0(x,y)\<=0\}\subset\IP^2_x{\times}\IP^1_y$ is equivalent to requiring $(x_0,x_1,x_2)\<=(0,0,0)$, which does not exist in $\IP^2_x$.
 With the restrictions~\eqref{e:fd12}, the $\d_i$ cannot be absorbed into $x_1,\cdots,x_6$ by their rescaling; the $\d_i$ may be thought of as parametrizing a small deformation.

Having introduced rational-monomial deformations, the putative pole-locus must be addressed, to which end Ref.~\cite{rBH-gB} employed the L'Hopital-like ``intrinsic limit'':
\begin{defn}[Ref.~\cite{rBH-gB}]\label{D:iLim}
Away from $\{x_5{=}0\}\<\cup\{x_6{=}0\}$, the zero locus of
 $f_{\!L\vec\d\,}(x)$ in~\eqref{e:fd12} is well defined, and allows solving, e.g., for $x_2\<=x_2(x_1,x_5,x_6)$:
\begin{equation}
  x_5\<\neq0:\quad
  f_{\!L\vec\d\,}(x)\<=0\quad\To\quad
  x_2^{~2} = -\frac{x_1^{~2} x_5 x_6 (x_5^{~5}+x_6^{~5})}{\d_1x_6+\d_2x_5}.
 \label{e:cond}
\end{equation}
Taking the $x_5,x_6\to0$ limits while restricting $x_2^{~2}$ to hold the \eqref{e:cond}-values preserves the vanishing of $f_{\!L\vec\d\,}(x)$, defines (a closure/completion of) the zero locus, $Z_{f_{\!L\vec\d\,}}$,
and is called the {\bfseries intrinsic limit}.
\end{defn}
\begin{remk}\label{R:iLimFinite}
The pole-locus in~\eqref{e:fd12} is $\sP\<\coeq\{x_5{=}0\}\<\cup\{x_6{=}0\}$, and the constraining condition~\eqref{e:cond} of Definition~\ref{D:iLim} is clearly well defined except along the zero locus of its denominator. That however intersects $\sP$ only at $\sP^\sharp\<\coeq\{x_5{=}0\}\<\cap\{x_6{=}0\}$, where L'Hopital's rule again gives~\eqref{e:cond} a well-defined value (zero). In general, such as when the requisite Laurent deformation has putative pole locations of higher order, the ``intrinsic limit'' will require additional iterations of applying L'Hopital's rule. Although the pole locations will have a growing complexity and order in higher dimensions and higher ``twist'' ($m$) in $\FF{m}$, it would seem that the ``intrinsic limit'' resolution of the ambiguity in defining the zero locus, $Z_{f}$, i.e., specifying its closure is a well defined procedure with a guaranteed finite completion. However, I am not aware of a proof.
\end{remk}
Constrained (conditional) limits are far from novel in physics in general, making the ``intrinsic limit'' of~\cite{rBH-gB} the ``obvious'' {\em\/physicsy\/} resolution of the ambiguity in defining the zero locus 
$Z_{f_{\!L\vec\d\,}}$; we return to this in \SS\,\ref{s:CC}.
 However, the use of the (constrained, moreover) {\em\/limiting procedure\/} definition the so-defined (closure of the) zero locus clearly veers outside the usual framework of algebraic geometry.
\begin{conj}\label{C:NonAlg}
 Laurent-deformed Calabi--Yau hypersurfaces closed/completed by the ``intrinsic limit'' (Definition~\ref{D:iLim}) are not algebraic varieties; they are toric spaces, equipped with a maximal $U(1)^n$-action, and corresponding $U(1)^n$- or even fully $U(1;\IC)^n$-equivariant (co)homology.
\end{conj}
Having veered outside the standard framework of complex algebraic (toric) geometry in trying to smooth the ``unsmoothable'' Tyurin-degenerate models opens a whole host of questions. For example, consider the preimage of $Z_{q_0}$, under the sequence of deformations, all the way in the left-hand side of the diagram~\eqref{e:DDDefo}, where it has many (complex-algebraic, regular-monomial) smoothing deformations. Since $Z_{q_0}\subset\FF[4]{(5000)}$ can be smoothed by Laurent (rational-monomial) deformations, it is not unreasonable to propose:
\begin{conj}\label{C:DefoLaurent}
The rational sections, $\vec\d q_0$, that smooth the Tyurin degenerate $Z_{q_0}$ are (in the diagram~\eqref{e:DDDefo}, rightward) deformation limits of regular (not rational-monomial) sections of the anticanonical bundles sufficiently (in the diagram~\eqref{e:DDDefo}, leftward) away from 
$\cK^*(\FF[4]{(5000)})$.
\end{conj}

\section{Transposition Mirrors}
\label{s:TM}
The inclusion of spaces the construction of which veers outside of the routine complex-algebraic (toric) geometry to include spaces that are not algebraic (such as the ``intrinsic limit''-completed/closed zero loci of Laurent defining polynomials) has rather novel consequences via mirror symmetry.

\subsection{Transposition as Multitope Swapping}
\label{s:TM1}
Mirror symmetry was discovered over a third of a century ago~\cite{rGP1}, considering so-called Fermat defining equations,
 $\big\{f_F(x)\<=\sum_i x_i^{~p_i}\<=0\big\}\subset\IP^4_{\vec w}$, where the weights $\vec w$ of the quasi-projective space are chosen so that each summand has the same total weight,
  $p_iw_i=p_jw_j$ (no summation), for all $i,j=0,\cdots,4$.
In a little over two years' time, this was extended to the fifteen other types of transverse defining equations with five monomials (the same as the number of variables, $x_i$),
 $\big\{f_e(x)\<=\sum_j \prod_i x_i^{~e_{ij}}\<=0\big\}$, the mirror of which is found as a particular finite quotient of the zero locus of the ``transpose polynomial,''
 $\big\{f_e^\sfT(y)\<\coeq\sum_j \prod_i y_i^{~e_{ji}}\<=0\big\}$, found by transposing the matrix of exponents, $e_{ij}\to e_{ji}$~\cite{rBH}. Less than two years later, a vast generalization to all (complex-algebraic) toric hypersurfaces was found~\cite{rBaty01}, readily applicable to GLSMs, and closely related to the earlier notion of ``transposition,'' which then will be used herein as a common descriptive identifier.

\begin{figure}[htb]
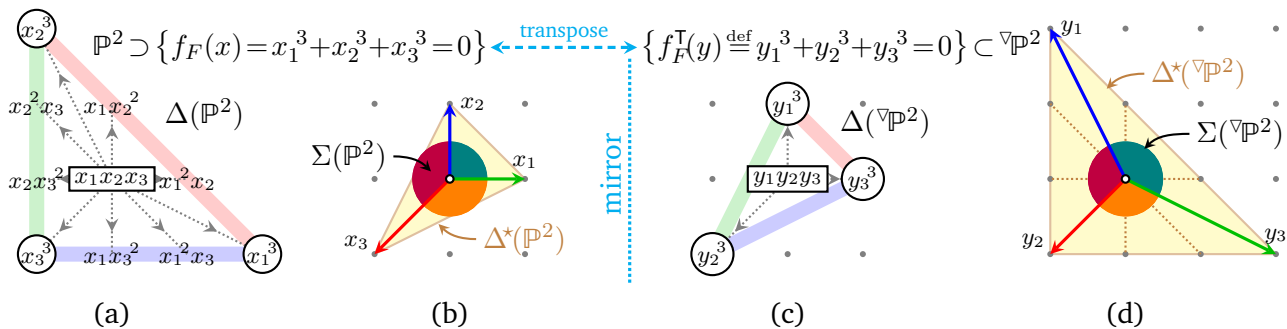

\raggedleft
  \TikZ{[thick]\path[use as bounding box](-1.5,-2)--(2.5,2.5);
    \path(35:1.5)node{$\D(\IP^2)$};
    \draw[red!20, line width=2mm](-1,2)--(2,-1);
    \draw[Green!20, line width=2mm](-1,2)--(-1,-1);
    \draw[blue!20, line width=2mm](-1,-1)--(2,-1);
    \draw[gray, densely dotted, midarrow=stealth](0,0)--(-1,2);
    \draw[gray, densely dotted, midarrow=stealth](0,0)--(-1,1);
    \draw[gray, densely dotted, midarrow=stealth](0,0)--(-1,0);
    \draw[gray, densely dotted, midarrow=stealth](0,0)--(-1,-1);
    \draw[gray, densely dotted, midarrow=stealth](0,0)--( 0,-1);
    \draw[gray, densely dotted, midarrow=stealth](0,0)--( 1,-1);
    \draw[gray, densely dotted, midarrow=stealth](0,0)--( 2,-1);
    \draw[gray, densely dotted, midarrow=stealth](0,0)--( 1, 0);
    \draw[gray, densely dotted, midarrow=stealth](0,0)--( 0, 1);
     \filldraw[fill=white](-1, 2)circle(2.5mm);
     \filldraw[fill=white](-1,-1)circle(2.5mm);
     \filldraw[fill=white]( 2,-1)circle(2.5mm);
    \path(-1, 2)node{\fnS$x_2\!^3$};
    \path(-1, 1)node{\fnS$x_2\!^2x_3$};
    \path(-1, 0)node{\fnS$x_2x_3\!^2$};
    \path(-1,-1)node{\fnS$x_3\!^3$};
    \path( 0, 1)node{\fnS$x_1x_2\!^2$};
    \path( 0,0)node[rectangle, fill=white, draw=black, inner sep=2pt]
        {\fnS$x_1x_2x_3$};
    \path( 0,-1)node{\fnS$x_1x_3\!^2$};
    \path( 1, 0)node{\fnS$x_1\!^2x_2$};
    \path( 1,-1)node{\fnS$x_1\!^2x_3$};
    \path( 2,-1)node{\fnS$x_1\!^3$};
    \path(-.4,1.75)node[right]
        {$\IP^2\<\supset
           \big\{f_F(x)\<=x_1^{~3}{+}x_2^{~3}{+}x_3^{~3}\<=0\big\}$};
    \path(0,-1.8)node{(a)};
            }
 \quad
  \TikZ{[thick]
    \path[use as bounding box](-1.5,-2)--(2.5,2.5);
    \filldraw[fill=yellow!25, draw=brown!50, line join=round]
        (1,0)--(0,1)--(-1,-1)--cycle;
    \foreach\x in{-1,...,1}\foreach\y in{-1,...,1}
     \fill[gray](\x,\y)circle(.4mm);
    \corner{(0,0)}{0}{90}{.5}{teal}
    \corner{(0,0)}{90}{225}{.5}{purple}
    \corner{(0,0)}{225}{360}{.5}{orange}
    \draw[very thick, Green, -stealth](0,0)--(1,0);
     \path(1,0)node[above]{\fnS$x_1$};
    \draw[very thick, blue, -stealth](0,0)--(0,1);
     \path(0,1)node[right]{\fnS$x_2$};
    \draw[very thick, red, -stealth](0,0)--(-1,-1);
     \path(-1,-1)node[above left=-2pt]{\fnS$x_3$};
    \filldraw[fill=white](0,0)circle(.5mm);
    \path(.3,-.8)node[brown, right]{$\D\!^{\star\!}(\IP^2)$};
    \draw[brown, -stealth](.35,-.8)to[out=180,in=315]++(-.5,.2);
    \path(-.7,.3)node[left]{$\S(\IP^2)$};
    \draw[-stealth](-.75,.3)to[out=0,in=135]++(.5,-.2);
    \path(0,-1.8)node{(b)};
            }
 \quad
  \TikZ{[thick]\path[use as bounding box](-1.5,-2)--(2.5,2.5);
    \path(30:1.5)node{$\D(\tP\IP^2)$};
    \draw[very thick, cyan, densely dotted](-2.1,1.6)--
        node[xshift=1pt, above=-1, rotate=90]{mirror}++(0,-3);
    \draw[very thick, cyan, densely dashed, stealth-stealth](-2,1.75)--
        node[above=-2pt]{\scriptsize transpose}++(-1.95,0);
    \foreach\x in{-1,...,1}\foreach\y in{-1,...,1}
     \fill[gray](\x,\y)circle(.4mm);
    \draw[red!20, line width=2mm](1,0)--(0,1);
    \draw[Green!20, line width=2mm](0,1)--(-1,-1);
    \draw[blue!20, line width=2mm](-1,-1)--(1,0);
    \draw[gray, densely dotted, midarrow=stealth](0,0)--(1,0);
    \draw[gray, densely dotted, midarrow=stealth](0,0)--(0,1);
    \draw[gray, densely dotted, midarrow=stealth](0,0)--(-1,-1);
    \filldraw[fill=white](1,0)circle(2.8mm);
    \filldraw[fill=white](0,1)circle(2.8mm);
    \filldraw[fill=white](-1,-1)circle(2.8mm);
    \path(1,0)node{\fnS$y_3\!^3$};
    \path(0,1)node{\fnS$y_1\!^3$};
    \path(-1,-1)node{\fnS$y_2\!^3$};
    \path( 0,0)node[rectangle, fill=white, draw=black, inner sep=2pt]
        {\fnS$y_1y_2y_3$};
    \path(-2.1,1.75)node[right]
        {$\big\{f_F^\sfT(y)\<\define 
                y_1^{~3}{+}y_2^{~3}{+}y_3^{~3}\<=0\big\}
            \<\subset\tP\IP^2$};
    \path(0,-1.8)node{(c)};
            }
 \quad
  \TikZ{[thick]
    \path[use as bounding box](-1.5,-2)--(2.5,2.5);
    \filldraw[fill=yellow!25, draw=brown!50, line join=round]
        (-1,2)--(-1,-1)--(2,-1)--cycle;
    \foreach\x in{-1,...,2}\foreach\y in{-1,...,2}
     \fill[gray](\x,\y)circle(.4mm);
    \draw[brown, densely dotted](0,0)--(-1,1);
    \draw[brown, densely dotted](0,0)--(-1,0);
    \draw[brown, densely dotted](0,0)--(0,-1);
    \draw[brown, densely dotted](0,0)--(1,-1);
    \draw[brown, densely dotted](0,0)--(1,0);
    \draw[brown, densely dotted](0,0)--(0,1);
    \corner{(0,0)}{-atan(1/2)}{{90+atan(1/2)}}{.5}{teal}
    \corner{(0,0)}{{90+atan(1/2)}}{225}{.5}{purple}
    \corner{(0,0)}{225}{{360-atan(1/2)}}{.5}{orange}
    \draw[very thick, Green, -stealth](0,0)--(2,-1);
     \path(2,-1)node[above]{\fnS$y_3$};
    \draw[very thick, blue, -stealth](0,0)--(-1,2);
     \path(-1,2)node[right]{\fnS$y_1$};
    \draw[very thick, red, -stealth](0,0)--(-1,-1);
     \path(-1,-1)node[above left=-2pt]{\fnS$y_2$};
    \filldraw[fill=white](0,0)circle(.5mm);
    \path(.2,1.4)node[brown, right]{$\D\!^{\star\!}(\tP\IP^2)$};
    \draw[brown, -stealth](.25,1.4)to[out=180,in=45]++(-.5,-.2);
    \path(.8,.6)node[right]{$\S(\tP\IP^2)$};
    \draw[-stealth](.85,.6)to[out=180,in=45]++(-.6,-.2);
    \path(0,-1.8)node{(d)};
            }
\caption{\label{f:P23tPm}
Combinatorial data for $\IP^2$ as a toric variety, and for $\tP\IP^2$:
 the Newton polygon $\D$ of (regular) anticanonical monomials (a), in Cox coordinates specified by the spanning polygon $\D\!^{\star}$ and fan $\S(\IP^2)$, (b). Similarly: Newton polygon $\D(\tP\IP^2)$ in (c) and spanning polygon $\D\!^{\star\!}(\tP\IP^2)$ and fan $\S(\tP\IP^2)$ in (d).}
\end{figure}

Following the above practice, consider for illustration a simple, 2-dimensional version of Greene--Plesser's 1990 construction~\cite{rGP1}, a Fermat cubic in $\IP^2$ (see Figure~\ref{f:P23tPm})
where the \eqref{e:F1F3N}-like fan was redrawn in Figure~\ref{f:P23tPm}\,(b) in the customary orientation and labeled by the Cox coordinates, so powers of $x_i$ in the Newton polytope, $\D(\IP^2)$ Figure~\ref{f:P23tPm}\,(a), grow in the direction of the $x_i$-arrow in $\S(\IP^2)$. This fan subdivides its ``spanning polytope,'' $\D\!^{\star\!}(\IP^2)$. The ``original'' Calabi--Yau 1-fold is simply the zero locus of the Fermat cubic Figure~\ref{f:P23tPm}\,(top left), $f_F(x)$. The Greene--Plesser mirror was then defined as a Calabi--Yau desingularization of a $\ZZ_3$-quotient of the {\em\/same\/} cubic in essentially the {\em\/same\/} $\IP^2$.

 In Figure~\ref{f:P23tPm}\,(b), this mirror quotient of the Fermat cubic is seen as the transpose (the matrix of exponents is self-transposed), combining monomials from the Newton polytope, $\D(\tP\IP^2)$ Figure~\ref{f:P23tPm}\,(c), given in terms of a distinct set of variables, $y_j$, specified by the fan of a distinct embedding space, $\S(\tP\IP^2)$, depicted in Figure~\ref{f:P23tPm}\,(d).
 As should be clear from the shapes in Figure~\ref{f:P23tPm},
 this implements mirror symmetry by swapping the roles of the spanning polytope, $\D\!^{\star\!}(X)$, and the Newton polytope, $\D(X)$. That is,
\begin{equation}
  \D(\tP X) \coeq \D\!^{\star\!}(X),
  \quad\text{and}\qquad
  \D\!^{\star\!}(\tP X) \coeq \D(X),
 \label{e:BSwap}
\end{equation}
where the ``original'' is an anticanonical hypersurface in $X$ and its mirror is a corresponding hypersurface in $\tP X$, which is specified by both $\D(\tP X)$ and $\D\!^{\star\!}(\tP X)$, as defined in~\eqref{e:BSwap}. This in fact {\em\/automagically\/} incorporates the requisite quotient desingularization, as seen after a brief toric pr\'ecis (see Refs.~\cite{rF-TV, rGE-CCAG, rCLS-TV, rM-MFans, Masuda:2000aa, rHM-MFs} for rigorous and complete details):
\begin{description}

\item[Cox coordinates:]
Cox coordinates of a toric variety $X$~\cite{rCox} are assigned to the vertices of $\D\!^{\star\!}(X)$.
Indeed, in Figure~\ref{f:P23tPm}\,(b), the homogeneous coordinates of
 $\IP^2$ are assigned to the vertices of $\D\!^{\star\!}(\IP^2)$.
On the mirrored right-hand side, the Cox coordinate of $\tP\IP^2$ are assigned to the vertices of $\D\!^{\star\!}(\tP\IP^2)\coeq\D(\IP^2)$ depicted in Figure~\ref{f:P23tPm}\,(d).

\item[Chart:]
A toric variety, $X$ is covered by an atlas of chars, each corresponding to a top-dimensional cone in the fan, $\S(X)$. The lattice {\em\/degree\/} of a cone, $\q$, is the dimension-rescaled volume of the 0-apex pyramid generated by the cone's lattice-primitive generators:
 $d(\q)\<\coeq(\dim[\q]!)\mathrm{Vol}[\triangleleft(\q)]$.

\item[MPCP-desingularization:]
Degree-$d$ cones encode $\IC^n/\ZZ_d$-like affine charts ($\IC^n/\ZZ_1\<\coeq\IC^n$), a {\em\/maximal projective crepant partial\/} (MPCP)-desingularization (blowup)~\cite{rBaty01} of which is encoded by a subdivision into degree-1 sub-cones; each generator cone of the subdivision encodes (part of) an exceptional locus of the MPCP-desingularization.

\item[Charts and gluing:]
A toric variety, $X$ is covered by an atlas of chars, each corresponding to a top-dimensional cone in the fan, $\S(X)$. Charts overlap where their (top-dimensional) cones have a common {\em\/facet\/} (codimension-1 cone in the boundary), which then specifies how the charts are glued together. The so-defined poset structure in the atlas of charts covering $X$ is in direct 1--1 correspondence to the poset of cones in the fan $\S(X)$.  A {\em\/complete\/} fan (where each codimension-1 facet adjoins two top-dimensional cones) corresponds to a compact toric space.

\item[Multifan layers:]
 In multifans (as used in Refs.~\cite{rBH-gB, Berglund:2022dgb, Berglund:2024zuz, Hubsch:2025sph}), a common facet $(k{-}1)$-cone, $\tau$, adjoins two $k$-cones, $\q,\q'$ that lie in distinct layers of a multifan that ``flip-folds'' at $\tau$, so that $\q\<\cap\q'\<=\tau$ contrary to appearances of a larger overlap. The hyperplane region spanned by the lattice-primitive generators of a cone is its (base) face in the spanning {\em\/multitope\/} of the multifan, which is assembled from the faces of all the cones and with the same poset structure; a precise correspondence with the rich (and varied) practices in the mathematical literature~\cite{rM-MFans, Masuda:2000aa, rHM-MFs, Masuda:2006aa, rHM-EG+MF, rH-EG+MFs2, Nishimura:2006vs, Ishida:2013ab, Davis:1991uz, Ishida:2013aa, buchstaber2014toric, Jang:2023aa, rK+T-pSympTM} remains to be determined.

\end{description}
Now compare $\S(\IP^2)$ in Figure~\ref{f:P23tPm}\,(b) with
 $\S(\tP\IP^2)$ in Figure~\ref{f:P23tPm}\,(d).
Each of the three top-dimensional cones of $\S(\IP^2)$ in Figure~\ref{f:P23tPm}\,(b) has degree 1~\cite{rBeast2}:
\begin{equation}
  d(\q_{12})=\det\bM{1&0\\[-1mm] \cline{1-2} 0&1}=1,\quad
  d(\q_{23})=\det\bM{~~0&~~1\\[-1mm] \cline{1-2}-1&-1}=1,\quad
  d(\q_{31})=\det\bM{-1&-1\\[-1mm] \cline{1-2}~~1&~~0}=1,
\end{equation}
so each encodes the familiar $\IC^2$-like chart, which jointly cover $\IP^2$. By contrast, each of the three top-dimensional cones of $\S(\tP\IP^2)$ in Figure~\ref{f:P23tPm}\,(d) has degree 3:
\begin{equation}
  d(\Q_{12})\<=\det\bM{~~2&-1\\[-1mm] \cline{1-2}-1&~~2}\<=3,~~
  d(\Q_{23})\<=\det\bM{-1&~~2\\[-1mm] \cline{1-2}-1&-1}\<=3,~~
  d(\Q_{31})\<=\det\bM{-1&-1\\[-1mm] \cline{1-2}~~2&-1}\<=3.
\end{equation}
They all encode identical $\IC^2/\ZZ_3$-like charts, which jointly cover 
 $\tP\IP^2$, thereby indicating the {\em\/global quotient,} $\IP^2/\ZZ_3$. Each chart requires two MPCP-desingularizing blowups, corresponding to the dotted lines in Figure~\ref{f:P23tPm}\,(d):
 $\tP\IP^2\<=\Bl^\uA_{\sss\text{MPCP}}[\IP^2/\ZZ_3]$.

This Figure~\ref{f:P23tPm}\,(d)-encoded quotienting and MPCP-desingularization in $D(\tP\IP^2)$ is ``inherited'' by the hypersurface 
 $\{f_F^\sfT(y)\<=0\}\<\subset\tP\IP^2$ --- where it {\em\/precisely\/} matches the complete Greene--Plesser prescription~\cite{rGP1}. Indeed, the Newton polytope $\D\!^{\star\!}(\tP\IP^2)$ consists {\em\/only\/} of monomials that are invariant under the\break
 $\ZZ_3:(y_1,y_2,y_3)\to(\w^2y_1,\w y_2,y_3)$ action.\footnote{This is precisely the analogue of the Greene--Plesser choice, and is obtained as follows: Start with a generic diagonal action
 $(x_1,x_2,x_3)\<\to(\a x_1,\b x_2,\g x_3)$, where the invariance of the individual cubes requires $\a^3\<=\b^3\<=\g^3\<=1$. The invariance of the fundamental monomial, $\Pi{x}$, (oft-quoted as ``the Calabi--Yau condition'') sets $\g=\a^2\b^2$. The overall projectivization subsumes the
 $\b\<=\a\<\neq1$ choice, $(\a,\a,\a)$,
leaving the equivalent choices:
 $(1,\a,\a^2)$, $(\a,1,\a^2)$ and $(\a,\a^2,1)$ with $\a\<\neq1$
as the remaining candidates.}
\begin{corl}\label{C:tPm-Q}
The zero locus of the transpose anticanonical section,
 $\{f^\sfT(y)\<=0\}\<\subset\tP{X}$, with the toric space $\tP{X}$ encoded by the fan that star-subdivides $\D(X)$ and $f^\sfT(y)$ by $\D\!^{\star\!}(X)$, requires neither further quotienting nor any desingularization: $\tP{X}$ is already {\em\/well prepared,} including all requisite MPCP-desingularization of all $\D(X)$-encoded local quotient singularities.
\end{corl}

\subsection{Simplicial Reductions}
\label{s:TM2}
\begin{figure}[htb]
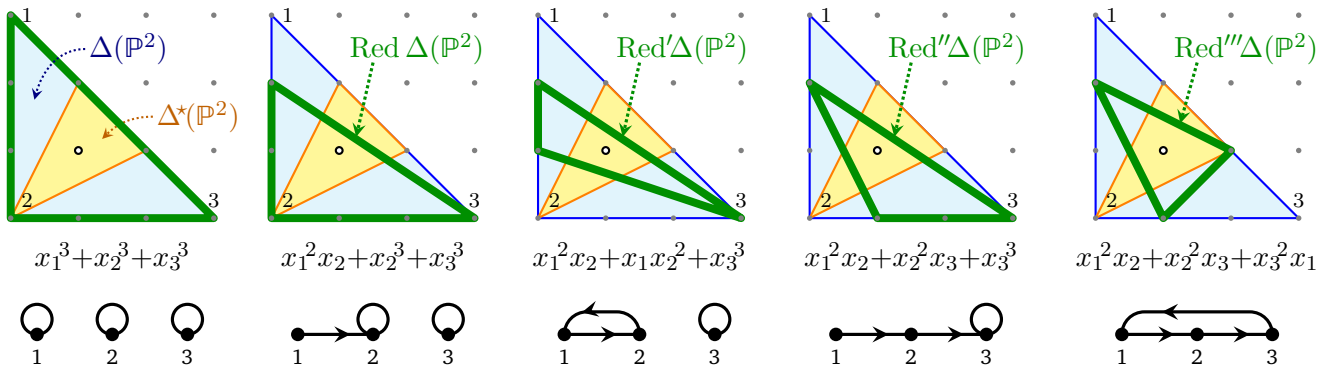

\raggedleft
$\begin{array}{@{}c@{\qquad}c@{\qquad}c@{\qquad}c@{\qquad}c@{}}
  \vC{\TikZ{[scale=.9, thick]
        \path[use as bounding box](-1,-1.3)--(2,2.2);
        \filldraw[fill=cyan!10, draw=blue, line join=round]
            (-1,2)--(-1,-1)--(2,-1)--cycle;
        \draw[fill=yellow!50, draw=orange, line join=round]
            (1,0)--(0,1)--(-1,-1)--cycle;
        \path(1,.5)node[right, orange!75!black]{$\D\!^{\star\!}(\IP^2)$};
         \draw[densely dotted, -stealth, orange!75!black]
            (1.1,.5)to[out=180,in=45]++(-.75,-.33);
        \draw[Green!75!black, line width=3pt, line join=round, opacity=.5]
            (-1,2)--(-1,-1)--(2,-1)--cycle;
        \foreach\x in{-1,...,2}\foreach\y in{-1,...,2}
         \fill[gray](\x,\y)circle(.4mm);
        \path(-1,2)node[right]{\scriptsize$1$};
        \path(-1,-1)node[above right]{\scriptsize$2$};
        \path(2,-1)node[above]{\scriptsize$3$};
        \path(0,1.5)node[right, Cobalt]{$\D(\IP^2)$};
         \draw[densely dotted, -stealth, Cobalt]
            (.1,1.5)to[out=180,in=75]++(-.75,-.75);
        \filldraw[fill=white, thick](0,0)circle(.5mm);
            }}
 &
  \vC{\TikZ{[scale=.9, thick]
        \path[use as bounding box](-1,-1.3)--(2,2.2);
        \filldraw[fill=cyan!10, draw=blue, line join=round]
            (-1,2)--(-1,-1)--(2,-1)--cycle;
        \filldraw[fill=yellow!50, draw=orange, line join=round]
            (1,0)--(0,1)--(-1,-1)--cycle;
        \draw[Green!75!black, line width=3pt, line join=round, opacity=.5]
            (-1,1)--(-1,-1)--(2,-1)--cycle;
        \foreach\x in{-1,...,2}\foreach\y in{-1,...,2}
         \fill[gray](\x,\y)circle(.4mm);
        \path(-1,2)node[right]{\scriptsize$1$};
        \path(-1,-1)node[above right]{\scriptsize$2$};
        \path(2,-1)node[above]{\scriptsize$3$};
        \path(0,1.5)node[right, Green!75!black]{$\Red\D(\IP^2)$};
         \draw[very thick, densely dotted, -stealth, Green!75!black]
            (.5,1.3)to++(-.25,-1.1);
        \filldraw[fill=white, thick](0,0)circle(.5mm);
            }}
 &
  \vC{\TikZ{[scale=.9, thick]
        \path[use as bounding box](-1,-1.3)--(2,2.2);
        \filldraw[fill=cyan!10, draw=blue, line join=round]
            (-1,2)--(-1,-1)--(2,-1)--cycle;
        \filldraw[fill=yellow!50, draw=orange, line join=round]
            (1,0)--(0,1)--(-1,-1)--cycle;
        \draw[Green!75!black, line width=3pt, line join=round, opacity=.5]
            (-1,1)--(-1,0)--(2,-1)--cycle;
        \foreach\x in{-1,...,2}\foreach\y in{-1,...,2}
         \fill[gray](\x,\y)circle(.4mm);
        \path(-1,2)node[right]{\scriptsize$1$};
        \path(-1,-1)node[above right]{\scriptsize$2$};
        \path(2,-1)node[above]{\scriptsize$3$};
        \path(0,1.5)node[right, Green!75!black]{$\Red'\!\!\D(\IP^2)$};
         \draw[very thick, densely dotted, -stealth, Green!75!black]
            (.5,1.3)to++(-.25,-1.1);
        \filldraw[fill=white, thick](0,0)circle(.5mm);
            }}
 &
  \vC{\TikZ{[scale=.9, thick]
        \path[use as bounding box](-1,-1.3)--(2,2.2);
        \filldraw[fill=cyan!10, draw=blue, line join=round]
            (-1,2)--(-1,-1)--(2,-1)--cycle;
        \filldraw[fill=yellow!50, draw=orange, line join=round]
            (1,0)--(0,1)--(-1,-1)--cycle;
        \draw[Green!75!black, line width=3pt, line join=round, opacity=.5]
            (-1,1)--(0,-1)--(2,-1)--cycle;
        \foreach\x in{-1,...,2}\foreach\y in{-1,...,2}
         \fill[gray](\x,\y)circle(.4mm);
        \path(-1,2)node[right]{\scriptsize$1$};
        \path(-1,-1)node[above right]{\scriptsize$2$};
        \path(2,-1)node[above]{\scriptsize$3$};
        \path(0,1.5)node[right, Green!75!black]{$\Red''\!\!\D(\IP^2)$};
         \draw[very thick, densely dotted, -stealth, Green!75!black]
            (.5,1.3)to++(-.25,-1.1);
        \filldraw[fill=white, thick](0,0)circle(.5mm);
            }}
 &
  \vC{\TikZ{[scale=.9, thick]
        \path[use as bounding box](-1,-1.3)--(2,2.2);
        \filldraw[fill=cyan!10, draw=blue, line join=round]
            (-1,2)--(-1,-1)--(2,-1)--cycle;
        \filldraw[fill=yellow!50, draw=orange, line join=round]
            (1,0)--(0,1)--(-1,-1)--cycle;
        \draw[Green!75!black, line width=3pt, line join=round, opacity=.5]
            (-1,1)--(0,-1)--(1,0)--cycle;
        \foreach\x in{-1,...,2}\foreach\y in{-1,...,2}
         \fill[gray](\x,\y)circle(.4mm);
        \path(-1,2)node[right]{\scriptsize$1$};
        \path(-1,-1)node[above right]{\scriptsize$2$};
        \path(2,-1)node[above]{\scriptsize$3$};
        \path(0,1.5)node[right, Green!75!black]{$\Red'''\!\!\D(\IP^2)$};
         \draw[very thick, densely dotted, -stealth, Green!75!black]
            (.5,1.3)to++(-.25,-.9);
        \filldraw[fill=white, thick](0,0)circle(.5mm);
            }}\\*[2mm]
    x_1\!^3 {+}x_2\!^3 {+}x_3\!^3
  & x_1\!^2x_2 {+}x_2\!^3 {+}x_3\!^3
  & x_1\!^2x_2 {+}x_1x_2\!^2 {+}x_3\!^3
  & x_1\!^2x_2 {+}x_2\!^2x_3 {+}x_3\!^3
  & x_1\!^2x_2 {+}x_2\!^2x_3 {+}x_3\!^2x_1\\*[0mm]
   \vC{\TikZ{\path[use as bounding box](0,-.5)--(2,.75);
               \foreach\x in{0,...,2}{\fill(\x,0)circle(.9mm);
                                      \draw[very thick](\x,.2)circle(2mm);}
               \foreach\x in{1,...,3}\path(\x-1,-.3)node{\scriptsize\x};
            }}
  &\vC{\TikZ{[very thick]\path[use as bounding box](0,-.5)--(2,.75);
               \foreach\x in{0,...,2}\fill(\x,0)circle(.9mm);
               \draw[midarrow=stealth](0,0)--(1,0);
               \draw(1,.2)circle(2mm);
               \draw(2,.2)circle(2mm);
               \foreach\x in{1,...,3}\path(\x-1,-.3)node{\scriptsize\x};
            }}
  &\vC{\TikZ{[very thick]\path[use as bounding box](0,-.5)--(2,.75);
               \foreach\x in{0,...,2}\fill(\x,0)circle(.9mm);
               \draw[midarrow=stealth](0,0)--(1,0);
               \draw[midarrow=stealth]
                   (1,0)to[out=90,in=0]++(-.3,.3)--++(-.4,0)to[out=180,in=90]++(-.3,-.3);
               \draw(2,.2)circle(2mm);
               \foreach\x in{1,...,3}\path(\x-1,-.3)node{\scriptsize\x};
            }}
  &\vC{\TikZ{[very thick]\path[use as bounding box](0,-.5)--(2,.75);
               \foreach\x in{0,...,2}\fill(\x,0)circle(.9mm);
               \draw[midarrow=stealth](0,0)--(1,0);
               \draw[midarrow=stealth](1,0)--(2,0);
               \draw(2,.2)circle(2mm);
               \foreach\x in{1,...,3}\path(\x-1,-.3)node{\scriptsize\x};
            }}
  &\vC{\TikZ{[very thick]\path[use as bounding box](0,-.5)--(2,.75);
               \foreach\x in{0,...,2}\fill(\x,0)circle(.9mm);
               \draw[midarrow=stealth](0,0)--(1,0);
               \draw[midarrow=stealth](1,0)--(2,0);
               \draw[midarrow=stealth]
                   (2,0)to[out=90,in=0]++(-.3,.3)--++(-1.4,0)to[out=180,in=90]++(-.3,-.3);
               \foreach\x in{1,...,3}\path(\x-1,-.3)node{\scriptsize\x};
            }}
  \end{array}$
\caption{\label{f:tvP23}
The distinct types of transverse cubics~{\cite{rAGZV-Sing1}}, classified as distinct simplicial 0-enclosing reductions of the complete Newton polytope}
\end{figure}

Generalizations of the Greene-Plesser construction to other types of transverse defining equations (with a minimal number of monomials) \'a~la~\cite{rBH} is then depicted by simplicial 0-enclosing reductions of the Newton polytope, depicted for $\IP^2$ (and dually, for $\tP\IP^2$) in Figure~\ref{f:tvP23},
where the (bigger, pale blue) Newton triangle, $\D(\IP^2)$, is depicted over the (smaller, yellow) triangle, $\D\!^{\star\!}(\IP^2)$. The thicker (green) triangles outline the $\Red\D(\IP^2)$ encode the defining polynomial underneath it, with the Ref.~\cite{rAGZV-Sing1, rBH}-styled depiction underneath that; Arnold's original classification~\cite{rAGZV-Sing1} guarantees completeness.

The combinatorially generated collection of different possible simplicial reductions evidently grows fast with the dimension and complexity of the considered polytope pairs, $\big(\Red_a\D\!^{\star\!}(X),\Red_b\D(X)\big)$.
Each of these specifies a mirror pair of Calabi--Yau manifolds. They are related by the fact that varying the reduction $\Red\D(X)$ corresponds to a deformation of the chosen defining equation, i.e., to a deformation of the complex structure, $Z_f\leadsto Z_{f'}$ --- both of which are mirrors to the ``same $Z_{f^\sfT}$,'' held ``fixed'' by not changing
 $\Red\D(\tP X)\<=\Red\D\!^{\star\!}(X)$.
 Since $\D(X)$ of the embedding space, $X$, of the ``original,'' $Z_f\<\subset X$ is the spanning polytope of the embedding space, $\tP X$ of the ``transposition-mirror,'' $Z_{f^\sfT}\<\subset\tP X$, varying $\Red\D(X)$ then also corresponds to varying the K\"ahler class (and also the symplectic structure~\cite{Kontsevich:1995wkA, Kontsevich:1994Mir}!) of $Z_{f^\sfT}\<\subset\tP X$, which thus are indeed ``held fixed'' as a complex manifold --- up to Bridgeland stability issues that make the choice of the complex and the K\"ahler structures subtly co-dependent~\cite{rD+G-DBrM}.
 In this sense,
\begin{corl}\label{C:MMM}
The original, unreduced polytope pair, $\big(\D\!^{\star\!}(X),\D(X)\big)$, may be regarded as encoding a ``generating machine'' for (``multiple'') mirror model pairs.
\end{corl}

\subsection{Flip-Folded Layers}
\label{s:TM3}
As the comparison of the two plots in Figure~\ref{f:F1F3M} indicates, extending the foregoing description of the transposition-mirror construction to hypersurfaces in $\FF{m}$ {\em\/defines\/} the embedding space,
 $\tP\FF{m}$, of the mirror model by the defining identities~\eqref{e:BSwap}. To this end, it is of key importance that the relation between the Newton and spanning polytope,
 such as $\D(\FF{m})\fif{\wtd}\D\!^{\star\!}(\FF{m})$,
is constructed using the (GLSM-deformation motivated) ``stripe''-wise computed transpolar operation --- rather than the algebraic-geometry standard, {\em\/global,\/} polar operation. This ``stripe''-wise {\em\/local\/} computation makes the transpolar operation expressly sensitive to all forms of non-convexity, such as evident in~(\ref{e:F1F3N},\,b) --- which the global computations in the standard polar operation obscure.
 
Abstracting from the plot in Figure~\ref{f:F1F3M}\,(b) and adapting the diagram~(\ref{e:F1F3N},\,b) to toric style of Figure~\ref{f:P23tPm} results in the double-duty diagrams in Figure~\ref{f:2F3tPm}.
\begin{figure}[htb]
\centering
  \TikZ{[thick]
      \path[use as bounding box](-1,-2.2)--(9,4.2);
      \foreach\x in{-1,...,1}\foreach\y in{-2,...,4}
       \fill[gray](\x,\y)circle(.4mm);
       \fill[cyan!20](0,0)--(1,-2)--(-1,4);
        \draw[blue, midarrow=stealth](1,-2)--(-1,4);
       \fill[red, opacity=.9](0,0)--(1,-2)--(1,-1);
       \draw[very thick, Rouge, midarrow=stealth](1,-1)--(1,-2);
      \filldraw[fill=white, draw=red](1,-2)circle(.6mm);
       \draw[very thick, purple!75!black, -stealth](0,0)--(1,-2);
       \fill[Green!21, opacity=.95](0,0)--(-1,4)--(-1,-1);
       \fill[brown!20, opacity=.95](0,0)--(-1,-1)--(1,-1);
       \draw[blue, midarrow=stealth](-1,4)--(-1,-1);
       \draw[blue, midarrow=stealth](-1,-1)--(1,-1);
      \draw[densely dotted](-1,3)--(0,0)--(-1,2);
      \draw[densely dotted](-1,1)--(0,0)--(-1,0);
      \draw[densely dotted](0,1)--(0,0)--(0,-1);
      \filldraw[fill=white](-1,4)circle(.6mm);
       \draw[very thick, teal!75!black, -stealth](0,0)--(-1,4);
      \filldraw[fill=white](-1,-1)circle(.6mm);
       \draw[very thick, Green!60!black, -stealth](0,0)--(-1,-1);
      \filldraw[fill=white, draw=red](1,-1)circle(.6mm);
       \draw[very thick, orange!75!black, -stealth](0,0)--(1,-1);
      \filldraw[fill=white](0,0)circle(.6mm);
      \path(-1,4)node[right, blue]{\fnS$y_1$};
      \path(-1,4)node[left]{\fnS$x_1^{~2}x_3^{~5}$};
      \path(-1,-1)node[below, blue]{\fnS$y_2$};
      \path(-1,-1)node[left]{\fnS$x_1^{~2}x_4^{~5}$};
      \path(1,-1)node[above, blue, xshift=1mm]{\fnS$y_3$};
      \path(1,-1)node[right, yshift=-1mm]{\fnS$x_2^{~2}\!/\!x_4$};
      \path(1,-2)node[left, blue]{\fnS$y_4$};
       \path(1,-1.6)node[right, Rouge]{\fnS$\Q_{34}$};
       \draw[ultra thick, densely dotted, Rouge, stealth-stealth]
          (1.6,-1.6)to[out=0, in=180]node[above=-2pt, rotate=38]
          {\fnS transpolar}++(3.85,1.62);
      \path(1,-2)node[right]{\fnS$x_2^{~2}\!/\!x_3$};
      \path(.3,.5)node[right]{$\D(\FF[2]3)\<=\D\!^{\star\!}(\tP\FF[2]3)$};
      \path(-.5,-1.8)node{(a)};
      \begin{scope}[xshift=7cm]
      \foreach\x in{-3,...,1}\foreach\y in{-1,...,1}
       \fill[gray](\x,\y)circle(.4mm);
      \fill[Green!60!brown!20](0,0)--(1,0)--(0,1);
      \fill[orange!75!brown!20](0,0)--(0,1)--(-1,0);
      \fill[purple!20](0,0)--(-1,0)--(-3,-1);
      \fill[teal!20](0,0)--(-3,-1)--(1,0);
      \draw[blue, midarrow=stealth](-3,-1)--(1,0);
      \draw[blue, midarrow=stealth](1,0)--(0,1);
      \draw[Rouge, midarrow=stealth](0,1)--(-1,0);
      \draw[Rouge, midarrow=stealth](-1,0)--(-3,-1);
      \filldraw[fill=white](1,0)circle(.6mm);
       \draw[very thick, Green, -stealth](0,0)--(1,0);
      \filldraw[fill=white](0,1)circle(.6mm);
       \draw[very thick, yellow!30!brown, -stealth](0,0)--(0,1);
      \filldraw[fill=white, draw=Rouge](-1,0)circle(.6mm);
       \draw[very thick, red, -stealth](0,0)--(-1,0);
      \filldraw[fill=white](-3,-1)circle(.6mm);
       \draw[very thick, blue, -stealth](0,0)--(-3,-1);
      \filldraw[fill=white](0,0)circle(.6mm);
      \path(-1,0)node[left=3pt, red]{\fnS$x_1$};
      \path(-1,0)node[above, blue, xshift=-2mm, yshift=1pt]
          {\fnS$y_1^{~2}y_2^{~2}$};
      \path(1,0)node[below]{\fnS$x_2$};
      \path(1,0)node[right, blue, yshift=1mm]{\fnS$y_3^{~2}y_4^{~2}$};
      \path(0,1)node[left]{\fnS$x_3$};
      \path(0,1)node[right, blue]{\fnS$y_1^{~5}\!/\!y_3$};
      \path(-3,-1)node[above]{\fnS$x_4$};
      \path(-3,-1)node[below, blue]{\fnS$y_2^{~5}\!/\!y_4$};
      \path(-.5,-1.5)node{$\D\!^{\star\!}(\FF[2]3)\<=\D(\tP\FF[2]3)$};
      \path(.5,-.8)node{(b)};
      \path(-5,3)node[right]{$\Bigg\{\begin{array}{@{}r@{\,=\,}l}
          f(x) & a_1 x_1^{~2}x_3^{~5} +a_2 x_1^{~2}x_4^{~5}
                +a_3\frac{x_2^{~2}}{x_4} +a_4\frac{x_2^{~2}}{x_3}\\[2mm]
          f^\sfT(x) &b_1 y_1^{~2}y_2^{~2} +b_1 y_3^{~2}y_4^{~2}
                +b_3\frac{y_1^{~5}}{y_3} +b_4\frac{y_2^{~5}}{y_4}
                      \end{array}$};
      \end{scope}
            }
\caption{\label{f:2F3tPm}
Combinatorial data for $\FF[2]3$ and $\tP\FF[2]3$:
 $\D(\FF[2]3)\<=\D\!^{\star\!}(\tP\FF[2]3)$ at left and
 $\D\!^{\star\!}(\FF[2]3)\<=\D(\tP\FF[2]3)$ at right;
 the transpose pair of ``cornerstone'' defining polynomials (top right)}
\end{figure}  
In Figure~\ref{f:2F3tPm}, $f(x)$ and $f^\sfT(y)$ are limited to the ``cornerstone'' monomials corresponding to the vertices of $\D(\FF[2]3)$ and $\D(\tP\FF[2]3)$, respectively. The matrix of exponents of one is the transpose of that of the other, and they are transverse polynomials provided
 $a_1a_3^5\neq a_2 a_4^5$ for $f(x)$, and
 $16 b_1^{~5}\neq -3125 b_2 b_3^{~2} b_4^{~2}$ and $b_2\neq 0$ for $f^\sfT(y)$. Relying on Corollary~\ref{C:tPm-Q}, this produces a mirror-pair of zero loci:
\begin{equation}
   \FF[2]3 \supset \big\{ f_(x)\<=0 \big\}
   \fif{~\text{mirror}~}
   \big\{ f^\sfT(y)\<=0 \big\} \subset \tP\FF[2]3.
\end{equation}

It remains however to better understand the toric space corresponding to $\tP\FF[2]3$. To this end, a few observations are in order, which are easy to generalize for all $m\<\geqslant3$:
\begin{enumerate}[itemsep=0pt, topsep=1pt, labelsep=1.17pc]

 \item The dotted lines in the $\D(\FF[2]m)$ diagram indicate the MPCP-desingularizations, so $\tP\FF[2]m$ is a smooth manifold, but it does not stem from a global finite quotient: Corresponding to the the four big (vertex-generated) cones, the four distinct charts (starting from the top-left vertex) correspond to the distinct MPCP desingularizations:
\begin{equation}
   \cU_{12}\<\approx\Bl^\uA_{m+1}[\IP^2/\ZZ_{m+2}],~~
   \cU_{23}\<\approx\Bl^\uA_1[\IP^2/\ZZ_2],~~
   \cU_{34}\<\approx\Bl^\uA_{m-3}[\IP^2],~~
   \cU_{41}\<\approx\Bl^\uA_1[\IP^2/\ZZ_2].
\end{equation}

 \item $\tP\FF[2]m$ has a maximal toric $\IC^2$-action and a \eqref{e:X1-6}-like GLSM specification:
\begin{equation}
    \begin{array}{r|r|cccc@{\,}l}
  &y_0 &y_1 &y_2 &y_3 &y_4 \\[-1pt] \cmidrule[.8pt]{1-6}
 \Tw{Q}^1 &-2(m{+}2) &m{-}2 & 0    & m &-2 &
    \multirow2*{~\smash{$\bigg\}U(1)^2$}}\\ 
 \Tw{Q}^2 &-2(m{+}2) & 0    &m{-}2 & m &-2 & \\ 
  \end{array}
 \label{e:y1-6}
\end{equation}
All other linear combinations with integral charges are, for $m\<\geqslant3$, non-negative linear combinations of $\Tw{Q}^1$ and $\Tw{Q}^2$, which are thereby the Mori vectors.

 \item The gluing
 $\cU_{23}\<{\#_{\m_3}}\cU_{34}$ and $\cU_{34}\<{\#_{\m_4}}\cU_{41}$
 is however nonstandard: In the Figure~\ref{f:2F3tPm}\,(a) depiction, the corresponding cones $\Q_{23}$ and $\Q_{34}$ (as well as $\Q_{34}$ and $\Q_{41}$) seem to partially overlap. This is a hallmark of {\em\/multifans\/}~\cite{rM-MFans, Masuda:2000aa, rHM-MFs}, which in general correspond to {\em\/torus manifolds,} and which are not algebraic varieties unless the multifan is in fact a (flat) fan.
 
 \item The multifan defined by the collection of central cones over the facets (``stripes'' of anticanonical monomials that are independent of one of the Cox coordinates) is well defined as a poset if the cone $\Q_{34}$ is understood to {\em\/flip-fold,} into a (Riemann-sheet like) layer under
 $\Q_{23}$ and over the layer of $\Q_{41}$. This way,
 $\Q_{23}\<\cap\Q_{34}\<=\Q_3$ and $\Q_{34}\<\cap\Q_{41}\<=\Q_1$ are 1-cones, consistently extending the 
 ``Separation Lemma''~\cite{rGE-CCAG, rCLS-TV}.
 
 \item \label{i:orient}
 This flip-folded character of the multifan centrally spanned by the multi-layered multigonal object $\D(\FF[2]m)$, which it star-subdivides, is well encoded by the {\em\/continuous orientation\/} of the closed cycle of cones,
\begin{equation}
 \Q_{\oA{12}}\to \Q_{\oA{23}}\to \Q_{\oA{34}}\to \Q_{\oA{41}}\>
 \TikZ{[line width=.4pt]\path[use as bounding box](0,0);
       \draw[midarrow=>, ->]
           (0,.11)to[out=0, in=-90]++(.15,.15)to[out=90, in=0]++(-.15,.15)
            --++(-4.525,0)to[out=180, in=90]++(-.15,-.15)
            to[out=-90, in=180]++(.15,-.15)--++(.1,0);
            }
 \label{e:orient}
\end{equation}
This is a key characteristic of the ``generalized legal loops''~\cite{rP+RV-12} --- which in fact {\em\/are\/} the 2-dimensional so-called {\em\/VEX multitopes\/}~\cite{rBH-gB, Berglund:2022dgb, Berglund:2024zuz}, the latter defined so that the transpolar operation acts within their class and always as an involution. 
 Thereby, VEX multitopes and their (local) transpolar involution generalize (in a GLSM-motivated fashion) both 
 ({\small\bf1})~the (convex) {\em\/reflexive\/} polytopes~\cite{rBaty01} and their (standard, global) {\em\/polar\/} operation to non-convex, flip-folded and otherwise multilayered VEX multitopes, as well as
 ({\small\bf2})~the ``generalized legal loops'' and their (local) dual operation of Ref.~\cite{rP+RV-12} to all higher dimensions.

 \item  Multifans do not uniquely encode torus manifolds, but it is not known how the continuous orientation of multifans and multitopes (item\,\ref{i:orient}, above), such as depicted in Figure~\ref{f:2F3tPm}, correlates with various combinatorial data considered in the literature to more precisely specify the available choices amongst (unitary) torus manifolds.
\end{enumerate}
In general, unitary torus manifolds, multifans and related combinatorial data corresponding to multi-layered multihedral complexes~\cite{rM-MFans, Masuda:2000aa, rHM-MFs, Masuda:2006aa, rHM-EG+MF, rH-EG+MFs2, Nishimura:2006vs, Ishida:2013ab}; see also~\cite{Davis:1991uz, Ishida:2013aa, buchstaber2014toric, Jang:2023aa} not only do not correspond complex algebraic varieties, but need not admit even an almost complex structure. The relatively simple, flip-folded multifans such as the illustration in Figure~\ref{f:2F3tPm}\,(a) may well encode a ``wrong'' sort of a blowup: The blowup of $\IP^2$ at a (smooth) point is depicted in toric geometry by the right-hand side subdividing operation:
\begin{equation}
\vC{\TikZ{[thick]\path[use as bounding box](-1.5,-1)--(1.5,1);
           \foreach\x in{-1,...,1}\foreach\y in{-1,...,1}
            \fill[gray](\x,\y)circle(.4mm);
           \draw[blue](-1,-1)--(1,0);
           \fill[cyan!30](0,0)--(-1,-1)--(1,0);
           \fill[red!75, opacity=.9](0,0)--(1,0)--(1,-1);
           \draw[Rouge](1,-1)--(1,0);
           \draw[-stealth](0,0)--(1,0);
           \fill[Green!75, opacity=.9](0,0)--(1,-1)--(0,1);
           \fill[blue!30](0,0)--(0,1)--(-1,-1);
           \draw[blue](1,-1)--(0,1)--(-1,-1);
            \path(1,-1)node[orange, rotate=36]{$\bigstar$};
            \path(1.05,-1.15)node[right, rotate=36]{$\star$};
           \draw[Rouge, densely dashed, ultra thick, -stealth]
               (0,0)--(-45:1.39);
           \draw[-stealth](0,0)--(0,1);
           \draw[-stealth](0,0)--(-1,-1);
            \path(1,0)node[right]{\fnS$1$};
            \path(0,1)node[left]{\fnS$2$};
            \path(-1,-1)node[left]{\fnS$3$};
           \filldraw[fill=white](0,0)circle(.5mm);
           \path(-.3,.6)node[left]{$\D\!^{\star\!}(\fE_1)$};
           \path(.33,-.9)node{(a)};
            }}
\quad\overset{\ttt\bf?}{\reflectbox{\Large$\leadsto$}}\quad
\vC{\TikZ{[thick]\path[use as bounding box](-1.5,-1)--(1.3,1);
           \foreach\x in{-1,...,1}\foreach\y in{-1,...,1}
            \fill[gray](\x,\y)circle(.4mm);
           \fill[red!30](0,0)--(1,0)--(0,1);
           \fill[blue!30](0,0)--(0,1)--(-1,-1);
           \fill[cyan!30](0,0)--(-1,-1)--(1,0);
           \draw[blue](1,0)--(0,1)--(-1,-1)--cycle;
           \draw[-stealth](0,0)--(1,0);
           \draw[-stealth](0,0)--(0,1);
           \draw[-stealth](0,0)--(-1,-1);
            \path(1,0)node[right]{\fnS$1$};
            \path(0,1)node[left]{\fnS$2$};
            \path(-1,-1)node[left]{\fnS$3$};
           \filldraw[fill=white](0,0)circle(.5mm);
           \path(-.3,.6)node[left]{$\D\!^{\star\!}(\IP^2)$};
           \path(.33,-.9)node{(b)};
            }}
\quad\hbox{\Large$\leadsto$}\qquad
\vC{\TikZ{[thick]\path[use as bounding box](-1.5,-1)--(1.5,1);
           \foreach\x in{-1,...,1}\foreach\y in{-1,...,1}
            \fill[gray](\x,\y)circle(.4mm);
           \fill[red!30](0,0)--(1,0)--(1,1);
           \fill[Green!30](0,0)--(1,1)--(0,1);
           \fill[blue!30](0,0)--(0,1)--(-1,-1);
           \fill[cyan!30](0,0)--(-1,-1)--(1,0);
           \draw[blue](1,0)--(1,1)--(0,1)--(-1,-1)--cycle;
           \draw[-stealth](0,0)--(1,0);
            \path(1,1)node[orange]{$\bigstar$};
            \path(1,1)node[right]{$\star$};
           \draw[Rouge, densely dashed, ultra thick, -stealth]
               (0,0)--(45:1.39);
           \draw[-stealth](0,0)--(0,1);
           \draw[-stealth](0,0)--(-1,-1);
            \path(1,0)node[right]{\fnS$1$};
            \path(0,1)node[left]{\fnS$2$};
            \path(-1,-1)node[left]{\fnS$3$};
           \filldraw[fill=white](0,0)circle(.5mm);
           \path(-.2,.6)node[left]{$\D\!^{\star\!}(\Bl^\uA_1[\IP^2])$};
           \path(.33,-.9)node{(c)};
            }}
 \label{e:oBo}
\end{equation}
In this (standard) blowup, depicted in~(\ref{e:oBo},\,c), the subdividing generator in the fan/spanning polytope lies within the cone being subdivided, here $\q_{12}$. Flip-folding this into a multifan/multitope depicted~(\ref{e:oBo},\,a) of the kind seen in Figure~\ref{f:2F3tPm}\,(a) effectively moves the subdividing generator,
from ``$\star$'' at $(1,1)$ to ``\rotatebox{36}{$\star$}'' at $(1,-1)$ --- which is {\em\/outside\/} the cone $\q_{12}$ that was being subdivided, and so might be called a {\em\/blow\2{out}.}
 The 1-point blowup of $\IP^2$ may also be described as a connected sum $\IP^2\<\#\overline{\IP^2}$, where the orientation of the second copy of $\IP^2$ has been reversed~\cite{Huybrechts:2005aa}. The toric diagram in~(\ref{e:oBo},\,a) then may well correspond to
 $\IP^2\<\#\IP^2$, which without this reversal of orientation on the second copy does not admit a global complex structure.\footnote{I am indebted to Prof.~M.~Masuda for suggesting this possibility.}
 This is supported by the computation of the self-intersection of the divisors corresponding to the star-labeled vertices~\cite{Berglund:2024zuz}:
 ({\small\bf1})~$[D_\star]^2\<={-}1$, as standard for the exceptional divisor in a standard blow-up;
 ({\small\bf2})~$[D_{\mkern-2mu\rotatebox{36}{\fnS$\star$}}]^2\<={+}1$,
which is opposite from the exceptional divisor in a standard blow-up.

 This mismatch in orientations causes the complex structure of the rest of $\IP^2$ to degenerate in $\fE_1$ at the exceptional divisor (corresponding to the 1-cone ``$*$'') and its local neighborhood. The situation is then most suggestively analogous to the degeneration of the symplectic structure in so-called ``pre-symplectic manifolds,'' the description of which also uses flip-folded multigons~\cite{rK+T-pSympTM}. This correlation with symplectic geometry is of course also motivated by the general notion that mirror symmetry swaps it with complex geometry~\cite{Kontsevich:1995wkA, Kontsevich:1994Mir}, but but requires much more detailed analysis and comparison:
\begin{conj}\label{C:FF}
Given a Calabi--Yau hypersurface $Z_f\<\subset X$ in a non-Fano toric variety $X$ of which the spanning polytope, $\D\!^{\star\!}(X)$, is non-convex,
 ({\small\bfseries1})~the transposition mirror may be found as a (transposed) hypersurface, $Z_{f^\sfT}$, 
 in a toric space, $\tP{X}$, corresponding to a flip-folded spanning multitope,
 $\D\!^{\star\!}(\tP{X})\<=\big(\D\!^{\star\!}(\tP{X})\big)\!^\wtd\<=\D(X)$.
 ({\small\bfseries2})~The toric space $\tP{X}$ is pre-complex, as may well be the mirror Calabi--Yau hypersurface in it,\footnote{It {\em\/may\/} be possible to find a Calabi--Yau hypersurface that avoids intersecting the complex structure obstruction in $\tP{X}$; there is then no reason for $Z_{f^\sfT}\<\subset\tP{X}$ not to be complex.} $Z_{f^\sfT}\<\subset\tP{X}$, in that their complex structures degenerate at isolated locations corresponding to the flip-folded elements of the spanning multitope, 
 $\D\!^{\star\!}(\tP{X})$.
\end{conj}
\begin{remk}\label{R:defects}
The properly general class of Calabi--Yau spaces is rarely openly specified, but is in fact certain to include {\bfseries\/stratified pseudo-manifolds\/} and other ``defects''; see, e.g.,~\cite{rAGM04, Hubsch:2002st, rAR01, rB-IS+ST, McNamara:2019rup}. A corresponding (co)homology theory that is consistent with mirror symmetry and conifold/geometric transitions has been defined in Ref.~\cite{rB-IS+ST}; see also Refs.~\cite{Collins:2021Spe, Anderson:2022bpo, Friedman:2024zid, Picard:2025Cal} for more recent discussion. In the present context, a $U(1)^n$- or even fully $U(1;\IC)^n$-equivariant version seems to be required. The ``home'' for such a structure is presumed to require the derived categories of coherent sheaves and (special) Lagrangian submanifolds~\cite{rD+G-DBrM}.
\end{remk}

Let us conclude with the observation that the ``flip-foldedness'' is not ``undone'' by straightforward algebro-geometric means, such as a {\em\/blowdown,} i.e., the corresponding {\em\/removal\/} of one of the ``offending'' 1-cones while fixing all other features. For example, the flip-folded cone $\Q_{23}$ has $\Q_3$ (at $(1,-1)$) as one of its generators, which may be thought of as an (\ref{e:oBo},\,a)-styled ``outside'' subdivision of the cone $\sfa\!\big((0,-1),(1,-2)\big)$, and so a ``blowout'' as discussed below~\eqref{e:oBo}. Collapsing its exceptional divisor encoded by $\Q_3$ (while holding fixed all cones not adjacent to
$\Q_3$) converts the flip-folded $\D(\FF[2]3)$ into a flat (simple, single-layered) albeit non-convex polygon. However, its ``stripe''-wise locally computed transpolar is now flip-folded:
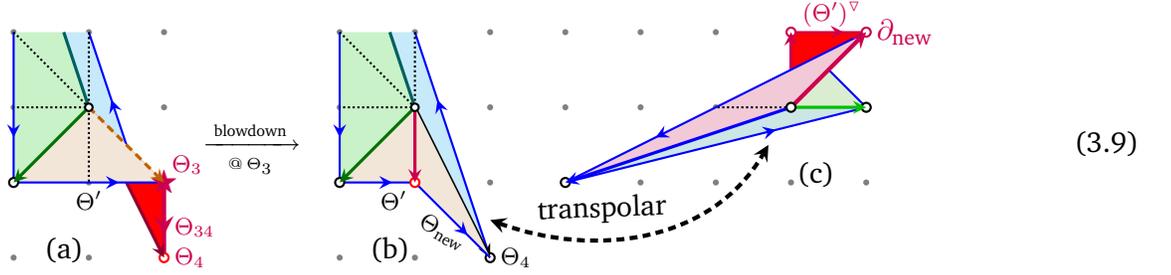
\begin{equation}
\vC{\TikZ{[thick]
      \path[use as bounding box](-1.2,-2.2)--(1.2,1.2);
      \foreach\x in{-1,...,1}\foreach\y in{-2,...,1}
       \fill[gray](\x,\y)circle(.4mm);
      \fill[cyan!20](0,0)--(1,-2)--(0,1)--(-1/3,1);
       \draw[blue, midarrow=stealth](1,-2)--(0,1);
      \fill[red, opacity=.9](0,0)--(1,-2)--(1,-1);
      \draw[ultra thick, Rouge, midarrow=stealth](1,-1)--(1,-2);
      \filldraw[fill=white, draw=red](1,-2)circle(.6mm);
       \draw[very thick, purple!75!black, -stealth](0,0)--(1,-2);
       \fill[Green!21, opacity=.95](0,0)--(-1/3,1)--(-1,1)--(-1,-1);
       \fill[brown!20, opacity=.95](0,0)--(-1,-1)--(1,-1);
       \draw[blue, midarrow=stealth](-1,1)--(-1,-1);
       \draw[blue, midarrow=stealth](-1,-1)--(1,-1);
      \draw[densely dotted](-1,1)--(0,0)--(-1,0);
      \draw[densely dotted](0,1)--(0,0)--(0,-1);
       \draw[very thick, teal!75!black](0,0)--(-1/3,1);
      \filldraw[fill=white](-1,-1)circle(.6mm);
       \draw[very thick, Green!60!black, -stealth](0,0)--(-1,-1);
      \path(0,-1)node[below=-1pt]{\fnS$\Q'$};
      \path[Rouge](1,-1)node{$\bigstar$};
      \path(1,-1)node[above right=-1pt, Rouge]{\fnS$\Q_3$};
      \path(1,-2)node[right, Rouge]{\fnS$\Q_4$};
      \path(1,-1.6)node[right, Rouge]{\fnS$\Q_{34}$};
       \draw[very thick, densely dashed, orange!75!black, -stealth]
           (0,0)--(1,-1);
      \filldraw[fill=white](0,0)circle(.5mm);
      \path(-.33,-1.9)node{(a)};
            }}
~~\xrightarrow[\sss @\,\Q_3]{\,\sss\text{blowdown}\,}~~
\vC{\TikZ{[thick]
      \path[use as bounding box](-1.2,-2.2)--(7,1.2);
      \foreach\x in{-1,...,1}\foreach\y in{-2,...,1}
       \fill[gray](\x,\y)circle(.4mm);
      \fill[cyan!20](0,0)--(1,-2)--(0,1)--(-1/3,1);
       \draw[blue, midarrow=stealth](1,-2)--(0,1);
      \filldraw[fill=white](1,-2)circle(.6mm);
       \draw[very thick, -stealth](0,0)--(1,-2);
       \fill[Green!21, opacity=.95](0,0)--(-1/3,1)--(-1,1)--(-1,-1);
       \fill[brown!20, opacity=.95](0,0)--(-1,-1)--(0,-1)--(1,-2);
       \draw[blue, midarrow=stealth](-1,1)--(-1,-1);
       \draw[blue, midarrow=stealth](-1,-1)--(0,-1);
       \draw[blue, midarrow=stealth](0,-1)--(1,-2);
      \draw[densely dotted](-1,1)--(0,0)--(-1,0);
      \draw[densely dotted](0,1)--(0,0);
       \draw[very thick, teal!75!black](0,0)--(-1/3,1);
      \filldraw[fill=white](-1,-1)circle(.6mm);
       \draw[very thick, Green!60!black, -stealth](0,0)--(-1,-1);
      \filldraw[fill=white, draw=red](0,-1)circle(.6mm);
       \draw[very thick, Rouge, -stealth](0,0)--(0,-1);
      \path(.5,-1.5)node[below=-2pt, rotate=-30]{\fnS$\Q_{\text{new}}$};
      \filldraw[fill=white](0,0)circle(.5mm);
      \path(0,-1)node[below left=-1pt]{\fnS$\Q'$};
      \path(1,-2)node[right]{\fnS$\Q_4$};
      \draw[ultra thick, densely dashed, stealth-stealth]
          (1,-1.5)to[out=-20, in=-120]
                  node[above, rotate=2, xshift=-6mm]{transpolar}++(3.7,1);
      \path(-.33,-1.9)node{(b)};
      \begin{scope}[xshift=5cm]
      \foreach\x in{-3,...,1}\foreach\y in{-1,...,1}
       \fill[gray](\x,\y)circle(.4mm);
      \fill[Green!60!brown!20](0,0)--(1,0)--(0,1);
       \draw[blue, midarrow=stealth](1,0)--(0,1);
      \fill[red, opacity=.85](0,0)--(0,1)--(1,1);
       \draw[Rouge, midarrow=stealth](0,1)--(1,1);
      \filldraw[fill=white, draw=Rouge](0,1)circle(.6mm);
       \draw[very thick, Rouge, -stealth](0,0)--(0,1);
     \fill[purple!21, opacity=.9](0,0)--(1,1)--(-3,-1);
      \fill[teal!20](0,0)--(-3,-1)--(1,0);
      \draw[densely dotted](0,0)--(-1,0);
      \draw[blue, midarrow=stealth](-3,-1)--(1,0);
      \draw[blue, midarrow=stealth](1,1)--(-3,-1);
      \path(.5,1)node[above=-2pt, Rouge]{\fnS~$(\Q')^\wtd$};
      \filldraw[fill=white](1,0)circle(.6mm);
       \draw[very thick, Green, -stealth](0,0)--(1,0);
      \filldraw[fill=white, draw=Rouge](1,1)circle(.6mm);
       \draw[ultra thick, Rouge, -stealth](0,0)--(1,1);
       \path(1,1)node[right, Rouge]{$\vd_{\text{new}}$};
      \filldraw[fill=white](-3,-1)circle(.6mm);
       \draw[very thick, blue, -stealth](0,0)--(-3,-1);
      \filldraw[fill=white](0,0)circle(.6mm);
      \path(.33,-.9)node{(c)};
      \end{scope}
            }}
 \label{e:FF}
\end{equation}
That is, unfolding $\D(\FF[2]3)$ ends up flip-folding $\D\!^{\star\!}(\FF[2]3)$ via their transpolar relationship:
\begin{corl}\label{C:nCtPFF}
 The transpolar image of a non-convex region is a flip-folded region, and {\em\/vice versa.} 
\end{corl}
 Here, the so-introduced edge,
 $\Theta_{\text{new}}\<=[(0,-1),(1,-2)]$,
encodes the pair of anticanonical monomials,
 $\{x_1x_2x_6\!^2,\, x_2\!^2\!/\!x_5\}$, that does not belong to any of the $\Pi{x}$-deformation defined ``stripes'' nor does it otherwise relate to the GLSM. Dually, the direction $\vd_{\text{new}}$ in~(\ref{e:FF},\,c) does not indicate any of the characteristic deformations of the fundamental monomial, $\Pi{x}$, in the plot of monomials in Figure~\ref{f:F1F3M}\,(b).
 The analogous blowdown at $\Q_4$ has identical consequences. Neither such {\em\/ad hoc,} by hand modification of the pair $(\D,\D\!^{\star\!})$ corresponds to $\FF[2]3$; they encode some different toric spaces.

The above-motivated need to extend {\em\/reflexive polytopes\/}~\cite{rBaty01} so as to include flip-folded and otherwise multilayered polyhedral objects (see also~\cite{Hibi:1995aa}) motivates a vast extension of the existing GLSM-motivated applications of toric geometry to:
\begin{defn}[Ref.~\cite{Berglund:2024zuz}]\label{D:VEX}
An $L$-lattice {\bfseries\/VEX multitope,} $\D\<\subset L_\IR\<\coeq(L\<{\otimes_\ZZ}\IR)$,
is a continuously orientable, possibly multi-layered, multihedral $n$-dimensional body with every facet at unit distance~\cite[Def.~4.1.4]{rBaty01} from $0\<\in L$,
 and is star-triangulated by a $0$-centered {\bfseries\/multifan}~\cite{rHM-MFs}:
 $\D\<\lat\S\<=(\s_\a,\pRec)$ is a facet-ordered poset of its $0$-centered cones,
 $\s_\a\<=\sfa(\q_\a\<\in\vd\D)$, were
 $\vs\<\pRec\s$ is a {\em\/facet\/}:
 $\vs\<\subset\vd\s$ and $\dim(\vs)\<=\dim(\s){-}1$.
 Then, $\D\<=(\q_\a,\pRec)$ has the same poset structure.
\end{defn}
\noindent
The toric geometry definition of ``unit distance'' perfectly corresponds to the $\Pi{x}$-deformation definition of ``distance-1.'' VEX multitopes were originally introduced~\cite{rBH-gB} in tandem with the transpolar operation (\SS\,\ref{s:FmFam1}), aiming to satisfy:
\begin{conj}[Ref.~\cite{Berglund:2024zuz}]\label{c:D**=D}
For an $L$-lattice VEX multitope, $\D$:
 ({\small\bfseries1})~$\D\!^\wtd\<\subset L^\vee_\IR$ is a VEX multitope,
and
 ({\small\bfseries2})~$(\D\!^\wtd)^\wtd\<=\D$: the transpolar operation
 (\SS\,\ref{s:FmFam1}) closes on VEX multitopes as an involution.
 ({\small\bfseries3})~The star-triangulation of $\D$ is {\em\/$L$-primitive}: lattice points of each star-triangulating simplex are only $\,0\<\in L$ and in $\vd\D$.
 If $\D$ has a unit-degree triangulation:
 ({\small\bfseries4})~$\D\!^\wtd$ has a corresponding unit-degree ($L^\vee$-)lattice triangulation,
    so both $\D$ and $\D\!^\wtd$ correspond to smooth toric spaces.
\end{conj}

\section{Concerns and Conclusions}
\label{s:CC}
The explicit $\vec\e$-varying deformations within the double deformation families of Calabi--Yau hypersurfaces in $\vec\e$-varying embedding spaces such as discussed in \SS\,\ref{s:FmFam3} insure that the so-constructed embedding spaces (here $\FF{m}$) are diffeomorphic to each other~\cite{rK-DefT}. These deformation families also include non-Fano embedding spaces and their ``unsmoothable'' (here Tyurin degenerate) Calabi--Yau hypersurfaces.

Our main concern here is whether there exists smoothings that preserve this diffeomorphism equivalence also at the level of the Calabi--Yau hypersurfaces (see~\cite{rReidK0, Friedman:1991} ) and the anticanonical bundles, $\cK^*(\FF{m})$. The Laurent deformations and their non-algebraic ``intrinsic limit'' completion/closure discussed herein would seem to satisfy this requirement, and is supported by a great deal of explicit computations, as reported earlier~\cite{rBH-gB, Berglund:2022dgb, Berglund:2024zuz, Hubsch:2025sph}. 

The main claims and statements are indicated throughout the foregoing discussion as
 Corollaries~\ref{C:GLSM-tP}, \ref{C:tPm-Q}, \ref{C:MMM}, and~\ref{C:nCtPFF}
and
 Remarks~\ref{R:GLSM-tP}, \ref{R:GLSM-CI}, \ref{R:RegRat}, \ref{R:WhyL}, \ref{R:chVars}, \ref{R:3} and~\ref{R:iLimFinite}. With regard to the smoothing of the ``unsmoothable'' Tyurin-degenerate models, these all refer to the non-algebraic (and presumably complex-structure degenerating) ``intrinsic limit.''
 The admittedly less completely justified but hopefully well-motivated statements are indicated as
 Conjectures~\ref{C:NonAlg}, \ref{C:DefoLaurent}, \ref{C:FF} and~\ref{c:D**=D}, followed in contrast by the decidedly algebro-geometric  Conjecture~\ref{C:Cox+} and ensuing questions, with which we close:

\paragraph{One More Thing: An Algebraic Alternative}
Contrasting the non-algebraic smoothing discussed above, let us close with a firmly algebro-geometric alternative: Ref.~\cite{Berglund:2022dgb} reports an $\FF[2]3$-generalization of a fractional mapping devised by D.~Cox for $\IP^2_{\smash{\sss(1:1:3)}}[5]$; this generalization converts the entire system of anticanonical sections of $\FF[2]3$ --- including the rational sections discussed here --- into the system of regular sections of a different algebraic variety. In fact, this can be generalized to all $\FF[2]m$: Start with the toric specification \'a la~\eqref{e:MI}, for $m\<\geqslant3$:
\begin{equation}
  f(x)= x_1^{~2} (x_3\<\oplus x_4)^{m+2} \oplus x_1\,x_2(x_3\<\oplus x_4)^2
   \oplus x_2^{~2} \Big(\frac1{x_3}\<\oplus \frac1{x_4}\Big)^{m-2}.
 \label{e:fx}
\end{equation}
The change of variables (selected to preserve the fibration nature of 
 $\FF[2]{m}$, from a very wide range of choices)
\begin{equation}
  (x_1,\, x_2,\, x_3,\, x_4)  \mapsto
  (z_0\sqrt{(w_0w_1)^{m-2}},\, z_1w_0^{~m}\sqrt{(w_0w_1)^{m-2}},\, w_0,\, w_1)
 \label{e:chCox}
\end{equation}
converts~\eqref{e:fx} into a completely regular polynomial:
\begin{equation}
 \begin{aligned}
  h(z)&= z_0\!^2\big(w_0^{~m-2}(w_0\<\oplus w_1)^{m+2}w_1^{~m-2}\big) \oplus
         z_0\,z_1\,\big(w_0^{~2m-2}(w_0\<\oplus w_1)^2\,w_1^{~m-2}\big)\\
      &\qquad \oplus
         z_1\!^2 \,w_0^{~2m}(w_0\<\oplus w_1)^{m-2},
 \end{aligned}
 \label{e:hz}
\end{equation}
which is a ({\em\/not completely\/}) generic $\cO\binom2{3m{-}2}$-section of
 $\IP^1_z{\times}\IP^1_w\<=\ssK[{r||c}{\IP^2&1\\ \IP^1&0}]\<=\FF[2]0$; for example, the $z_1\!^2$-term omits $(w_0\<\oplus w_1)^{2m-1}w_1^{~m-1}$ factors.
 Unsurprisingly, this fractional change of variables has a non-constant Jacobian, $\frac{\vd(x)}{\vd(z,w)}\<=w_0^{~m}(w_0w_1)^{m-2}$ and is ill-defined over the poles of $\IP^1_w$ --- very much like~\eqref{e:Segre7}. However, very much {\em\/unlike\/}~\eqref{e:Segre7}, the mapping~\eqref{e:chCox} relates the
 Laurent, ``intrinsic limit''-defined and transversal Calabi--Yau hypersurfaces $Z_f\<\subset\FF[2]3$ to
 regular, transverse hypersurfaces $Z_h\<\subset\FF[2]0$ {\em\/of general type\/}:
 $c_1(Z_h)=\binom{0}{\,4{-}3m}$, which is {\em\/negative\/} over $\IP^1_w$.
 Owing to the appearance of both roots and powers in~\eqref{e:chCox}, the relationship between $Z_f$ and $Z_h$ must involve both multiple covers and (presumably a suitable desingularization of) quotients. Any relationship between this ``regularizing'' change of variables~\eqref{e:chCox} and the geometry of hypersurfaces $Z_h$ of general type on one hand and the Laurent-deformed Calabi--Yau models discussed here would seem to be worthwhile: It is likely to provide new information about the novel, Laurent-deformed Calabi--Yau constructions, and perhaps a firmly algebraic reinterpretation of the ``intrinsic limit'' completion/closure of the Laurent-deformed hypersurfaces.

The relative ease (and the wide degree of freedom) in generalizing Cox's fractional mapping to the anticanonical system of all $\FF{m}$ motivates:
\begin{conj}[Regularizing]\label{C:Cox+}
All Calabi--Yau Laurent hypersurfaces encoded by VEX-multitopes of anticanonical sections are also describable via suitable fractional transformations in terms of {\bfseries general type} complete intersections of hypersurfaces in projective spaces,
 with strictly {\bfseries regular} defining polynomials, i.e., sums of monomials with only non-negative powers.
\end{conj}
By providing a strictly complex-algebraic bypass around the putative pole-singularity issue in Laurent hypersurfaces, fractional mappings such as~\eqref{e:fx}--\eqref{e:hz} raise obvious questions about the scope, differences and possible overlap between the distinct logical possibilities:\footnote{For string theory applications, this requires a (co)homology theory consistent with mirror symmetry and conifold/geometric transitions including various associated ``defects''~\cite{rAGM04, Hubsch:2002st, rB-IS+ST, McNamara:2019rup}, akin to the framework developed in Ref.~\cite{rB-IS+ST}, but with the GLSM $U(1)^n$- or even full $U(1;\IC)^n$-equivariance incorporated.}
 ({\small\bf1})~What is the (co)homological difference from the
    ``intrinsic limit''?
 ({\small\bf2})~Do they remain in the same deformation 
    class~\eqref{e:DDDefo}, or are there ``vanishing (or emerging) cycles''?
    That is, do the Betti or Hodge numbers differ, and how?
 ({\small\bf3})~How does the symplectic structure on these models (and homological Mirror symmetry) vary in the deformation process~\eqref{e:DDDefo}?
Eventually and if interested in string theory application, one would need to know: 
 ({\small\bf4})~Do the Weil--Petersson--Zamolodchikov metric-normalized Yukawa couplings (structure constants in the multiplicative $H^1(T)$ and $H^1(T^*)$ cohomology groups) differ, and how? 
--- but that is becoming a really tall order for now.

\paragraph{Acknowledgments:}
 First and foremost, I should like to thank Dr.~B.~Park for the kind invitation and opportunity to present this work.
 I am deeply grateful to Per Berglund for decades of collaborations including on the topics discussed here, and Corollary~\ref{C:tPm-Q} in particular,
 to Mikiya Masuda for his explanations of {\em\/unitary torus manifolds,}
 to Marco Aldi, Yong Cui and Amin Gholampour for insightful and instructive discussions.
 I am grateful to
 the Mathematics Department of the University of Maryland, and 
 the Physics Department of the University of Novi Sad, Serbia,
for recurring hospitality and resources.

\begingroup
\small\raggedright
\bibliographystyle{utphys}
\bibliography{RefsHR}
\endgroup

\end{document}